\newcommand{\ignore}[1]{}
\newcommand{\sem}[1]{\ensuremath{[\![#1]\!]}}
\newcommand{\angl}[1]{\langle#1\rangle}
\newcommand{\trans}[1]{\langle#1\rangle}
\newcommand{\DLT}{$\textsf{D\!T}_{\!\text{lin}}$\,}
\newcommand{\DBT}{$\textsf{D\!T}_{\!\text{uc}}$\,}
\newcommand{\DBTB}{$\textsf{D\!T}^{\text{a}}_{\!\text{uc}}$\,}
\newcommand{\DB}{$\textsf{D\!B}$\,}
\newcommand{\DLTB}{$\textsf{D\!T}^{\text{a}}_{\!\text{lin}}$\,}
\newcommand{\DBTR}{$\textsf{D\!T}^{\text{r}}_{\!\text{uc}}$\,}
\newcommand{\DBTI}{$\textsf{D\!T}^{\text{i}}_{\!\text{uc}}$\,}
\newcommand{\DLTI}{$\textsf{D\!T}^{\text{i}}_{\!\text{lin}}$\,}
\newcommand{\DLTR}{$\textsf{D\!T}_{\!\text{lin}}^{\text{r}}$\,}
\newcommand{\dom}{\textsf{dom}\,}
\newcommand{\C}{{\mathcal C}}
\renewcommand{\P}{{\mathcal P}}
\renewcommand{\L}{{\mathcal L}}

\newcommand{\G}{{\mathcal G}}
\newcommand{\T}{{\mathcal T}}

\newcommand{\In}{I}

\newcommand{\pref}{\textsf{pref}\,}
\newtheorem{example}{Example}
\newtheorem{lemma}{Lemma}
\newenvironment{proof}{\smallskip\noindent\textbf{Proof.}\;}{\qed\\[1ex]}
\newtheorem{theorem}{Theorem}
\newtheorem{corollary}{Corollary}
\documentclass[preprint,12pt]{elsarticle}

\usepackage{amssymb}
\usepackage{amsmath}

\journal{Nuclear Physics B}

\begin{document}

\begin{frontmatter}



\title{When is a Bottom-Up Deterministic Tree
Translation Top-Down Deterministic?}

\author[label1]{Sebastian Maneth}
\affiliation[label1]{%
    organization={Universit{\"a}t Bremen},
    addressline={Data Science Center, Bibliothekstra{\ss}e 5},
    city={Bremen},
    postcode={28359},
    state={},
    country={Germany},
    email={maneth@uni-bremen.de}
    }

\author[label2]{Helmut Seidl} 
\affiliation[label2]{organization={TU M\"unchen},
            addressline={School of CIT, Boltzmannstra{\ss}e 3}, 
            city={Garching},
            postcode={85748}, 
            state={},
            country={Germany},
	    email={helmut.seidl@tum.de}
	    }

\begin{abstract}
We consider two natural subclasses of deterministic top-down tree-to-tree transducers,
namely, linear and  uniform-copying transducers.
For both classes we show that it is decidable whether the translation of a transducer \emph{with}
look-ahead can be realized by a transducer 
from the same class 
\emph{without} look-ahead.
The transducers constructed in this way, may still make use of \emph{inspection},
i.e., have an additional tree automaton restricting the domain.
We provide a second procedure which decides whether inspection can be removed.
The procedure relies on a precise abstract interpretation of inspection requirements and
a dedicated earliest normal form for linear as well as uniform-copying transducers which can be 
constructed in polynomial time. As a consequence, equivalence of these
transducers can be decided in polynomial time.
Applying these results to deterministic bottom-up tree transducers, we obtain that it is
decidable whether or not their translations can be realized by deterministic linear or uniform-copying
top-down transducers without look-ahead (but with inspection) --- 
or without both look-ahead and inspection.
Look-ahead removal has been known to be a notoriously difficult problem.
To the best of our knowledge, this paper is the first to present look-ahead removal
for natural and known subclasses of top-down tree transducers.
\end{abstract}


\begin{highlights}
\item  	We show that it is decidable whether the translation of uniform-copying deterministic 
	top-down transducer with look-ahead can be realized by a transducer from the same class 
	without look-ahead but with inspection.

\item 	We show that it is decidable whether the translation of uniform-copying deterministic 
	top-down transducer without look-ahead, but with inspection can be realized by a transducer 
	from the same class but without inspection.
\end{highlights}



\end{frontmatter}




\section{Introduction}\label{s:intro}

Even though top-down and bottom-up tree transducers are well-studied formalisms
that were introduced already in the 1970's (by Rounds~\cite{DBLP:journals/mst/Rounds70} and 
Thatcher~\cite{DBLP:journals/jcss/Thatcher70} independently,
and by Thatcher~\cite{BU}, respectively), some fundamental questions 
have remained open until today.
A prominent example of such a question is:
can we decide for a given deterministic bottom-up tree transducer
whether or not its translation can be realized by a top-down tree transducer?
We answer this question affirmatively, however, for a slight restriction on the
considered top-down tree transducers: they must be \emph{uniform-copying}.
This means that all copies of the same input subtree must be processed by the same state.

It is well-known that for every deterministic bottom-up tree transducer an equivalent
deterministic top-down tree transducer can be constructed which, however, makes use of regular 
look-ahead~\cite{DBLP:journals/mst/Engelfriet77}.
That transducer indeed is \emph{uniform-copying}.
The question which we ask therefore is: can regular look-ahead in \emph{uniform-copying} transducers
be eliminated?
We prove that this is indeed the case: given a uniform-copying (or linear) top-down tree transducer
with look-ahead it is decidable whether or not it can be realized by a uniform-copying (or linear)
top-down tree transducer \emph{without} look-ahead.
Note that is is well-known that equivalence is decidable for all of these
deterministic tree transducers~\cite{DBLP:journals/jcss/EngelfrietMS09}.
Hence, an equivalent transducer in the desired class can be found (by brute force),
once it is known to exist. We present more direct constructions for each case when we decide one
class of transducers within a larger class (but we refrain from detailed complexity analysis
of the running time of our constructions).

In order to answer the question,
we provide a canonical earliest normal form for \emph{uniform-copying} (as well as for linear) 
deterministic top-down transducers with and without look-ahead.
We prove that if an earliest such transducer $A$ can be realized by such
a transducer $A'$ \emph{without} look-ahead (but with input inspection), then 
$A$ must be 
synchronizing,
twinning, and
erasing.
To understand the \emph{synchronizing} property, consider 
a transducer with look-ahead that translates an input tree of the form
\[
a(b(f(t_1,t_2)))\text{ into the tree }\langle a,b\rangle(f(t_1,t_2)),
\]
where
$\langle a,b\rangle$ is a binary output symbol.
Clearly this translation \emph{can} be done by a transducer without look-ahead:
it outputs nothing at the root node, keeps the node label in its state,
and at its child outputs the corresponding binary symbol. 
Now consider that
\[
a(b(f(g(t_1,t_2),t_3)))\text{ is translated to }
\langle a,b,g\rangle(f(g(t_1,t_2),t_3)),
\]
where an $a$ could be a $b$ (or vice versa) and an $f$ could be a $g$ (or vice versa).
Such a translation can\emph{not} be realized by a transducer without look-ahead.
The information about the $g$-node cannot be ``synchronized'' at the $f$-node
because it comes after the output must be produced
(contradicting the order of origins of output nodes~\cite{DBLP:conf/pods/LemayMN10}, 
see also~\cite{DBLP:journals/iandc/FiliotMRT18,DBLP:journals/jcss/FulopM16}).

The \emph{twinning} property is similar to the string case~\cite{DBLP:journals/tcs/Choffrut77} 
(see also~\cite{DBLP:journals/tcs/BealC02,DBLP:books/lib/Berstel79}), but now for paths. 
To understand the \emph{erasing} property, consider a transducer
with the following rules.
\[
{
\begin{array}{lcllcl}
q_0(a(x_1:h_e))&\to& a(a(e))&
q_{\text{id}}(a(x_1: h_f))&\to&a(q_{\text{id}}(x_1))\\
q_0(a(x_1:h_f))&\to& a(q_{\text{id}}(x_1))\ \qquad\qquad&
q_{\text{id}}(f)&\to&f\\
q_0(f)&\to& f
\end{array}}
\]
Here, input trees are of the form $a(\cdots a(e)\cdots)$ or $a(\cdots a(f)\cdots)$.
The look-ahead automaton has two states $h_e$ and $h_f$, indicating that the input tree
is of the first form ($e$-leaf) or the second form ($f$-leaf).
The transducer translates input trees of the first form to the fixed tree $a(a(e))$ and
realizes the identity on trees of the second form.  
This translation can\emph{not} be realized by a transducer without look-ahead.
The \emph{erasing} property demands that if an input path depends on two different 
look-ahead states $h_1,h_2$, where for $h_1$ a constant output tree is produced
(viz. the tree $a(a(e))$), then for $h_2$ no output may be produced in any loop.

Given a \emph{uniform-copying}
transducer $A$ with look-ahead that is synchronizing, twinning, and erasing we construct
an equivalent \emph{uniform-copying} transducer with \emph{inspection} (if it exists), i.e., a
\emph{uniform-copying} transducer where the domain is given separately via some 
top-down deterministic tree automaton.
%

The \emph{third highlight} of our contribution is a procedure that removes inspection (if possible).
The idea here is quite different from what we have discussed until now.
Let us consider an example. The domain automaton accepts trees of
the form $f(t_1,t_2)$ where $t_2$ is an arbitrary binary tree (with internal nodes
labeled $f$ and leaves labeled $a$ or $b$) and $t_1$ is a tree which has a left-most
leaf labeled $a$ and a right-most leaf labeled $b$.
The transducer has this rule:
\[
{
q_0(f(x_1,x_2))\to f(f(b,b),q_{\text{id}}(x_2))},
\]
where state $q_{\text{id}}$ realizes the identity.
Does there exist an equivalent top-down tree transducer \emph{without} inspection?
As it turns out, the answer is ``yes''. The idea is that the output subtree
$f(b,b)$ can be used to simulate inspection!
These are the rules of an equivalent transducer without inspection:
\[
{
\begin{array}{lll@{\quad}lll}
q_0(f(x_1,x_2))&\to& f(q(x_1),q_{\text{id}}(x_2))	&
q(f(x_1,x_2))&\to& f(q_a(x_1),q_b(x_2))			\\
q_a(f(x_1,x_2))&\to& q_a(x_1)				&
q_b(f(x_1,x_2))&\to& q_b(x_2)				\\
q_a(a)&\to& b 						&
q_b(b)&\to& b
\end{array}}
\]
We show that it is decidable for a given top-down deterministic tree language, whether
or not it can be simulated on a given output tree. The challenge now is that 
it may be necessary to \emph{delay} outputting certain output subtrees, until rules
are encountered which require these output trees for simulating their inspection needs.
Similar as before, such delay is only possible along input paths and must stop
when two input subtrees of an input node are processed.
Using a precise abstract interpretation of inspection needs we are able
to determine if sufficiently large output subtrees can be made available in order
to meet all required inspection needs. 
%
The approach we have sketched here is quite different from
earlier methods for look-ahead removal~\cite{DBLP:journals/tcs/EngelfrietMS16} 
that rely on \emph{difference bounds}, i.e., the 
differences in the translation with respect to different look-ahead states.
In order to obtain an effective construction, however, so far a variety of technical restrictions
had to be introduced --- implying that even for \emph{linear} deterministic top-down tree transducers,
look-ahead removal remained an open problem.
To the best of our knowledge, this paper  is the first to present look-ahead removal
for natural and known subclasses of top-down tree transducers.

Our result is an example of a \emph{definability problem}, i.e., given classes $X,Y$
of transductions, is it decidable for each $y\in Y$ whether or not an $x\in X$ exists such
that $x$ is equivalent to $y$. Many such problems have been considered in the past.
Fon instance, $X$ and $Y$ may be classes of languages, such as the regular and
the context-free languages, respectively. In this case, it is well-known by 
Greibach's Theorem~\cite{DBLP:journals/mst/Greibach68}
that the corresponding definability problem is \emph{undecidable}.
Interestingly, the situation changes if $Y$ is diminuished to the deterministic context-free
languages (i.e., those recognized by deterministic push-down automata). Then definability
is indeed decidable~\cite{DBLP:journals/jacm/Valiant75}.
In terms of \emph{tree languages}, such decidable definability
results are rare. One example is that for a given tree homomorphism $h$ and a regular tree langauge $R$
is is decidable whether or not $h(R)$ can be defined by a regular tree langauge~\cite{DBLP:journals/jacm/GodoyG13}.
This result was later generalized to a weighted setting~\cite{DBLP:conf/stacs/MalettiNP24}.

Transductions are somewhat more rigid than languages and thus better suited for decidable
definability problems. In terms of string transductions, an important decidable definability
problem was identified recently: for a given functional two-way transducer it is decidable
whether or not an equivalent one-way transducer 
exists~\cite{DBLP:conf/lics/FiliotGRS13,DBLP:journals/lmcs/BaschenisGMP18}.
In terms of tree transducers, this result has been extended to attributed tree transducers
with monadic output~\cite{DBLP:conf/wia/ManethV23}.
Recently, the regular string transductions have received much attention,
see, for instance~\cite{DBLP:journals/dagstuhl-reports/AlurBFMW23}.
It was shown that for a given regular string transducer, it is decidable whether its
translation has size increase $O(n^k)$, for any given $k$~\cite{DBLP:conf/lics/Bojanczyk23a}.
In terms of tree transducers, it is known whether or not a given composition of deterministic macro tree transducers
is of linear size increase~\cite{DBLP:journals/siamcomp/EngelfrietM03}; 
note that in the affirmative case, the translations is indeed
definable by an MSO transducer (Monadic Second-Order logic). Recently also linear height-increase
and linear size-to-height increase was shown to be decidable for a given macro tree
transducer~\cite{DBLP:conf/icalp/GallotM0P24}.

%
%
%
%

This paper is an extended version of~\cite{DBLP:conf/icalp/ManethS20}
which as been presented at ICALP'2020.

\section{Basics}\label{s:basics}

Let $\Sigma$ denote a ranked alphabet. Then $\Sigma_k$ is the set of all symbols in $\Sigma$ of
rank $k$. As usual, we define the set $\T_\Sigma$ of all (finite) \emph{trees} over $\Sigma$ 
as the set of all strings $t = f(t_1,\ldots,t_k)$ where $f\in\Sigma_k$ for some $k\geq 0$ and 
$t_1,\ldots,t_k\in\T_\Sigma$. 
For convenience, we also write $f$ for $f()$ if $f$ is of rank 0.
A subtree of that form is also called \emph{leaf}.
%
Let $X=\{x_i\mid i\in{\mathbb N}\}$ denote an infinite set
of distinct \emph{variables} which is disjoint from any other occurring ranked alphabet 
(be it the input or the output alphabet of a transducer). 
All elements of the set $X$ are assumed to have rank $0$.
For a finite set $J\subseteq{\mathbb N}$ we denote by $X_J$ the set
of variables $\{x_j\mid j\in J\}$,
and we write $\T_\Sigma(X_J)$ for the set of all trees $t$ 
over $\Sigma\cup X_J$
where each variable from $X_j$ occurs at least once.
E.g., $f(h(x_2),a)\in\T_\Sigma(\{x_2\})$ where
$f\in\Sigma_2$,
$h\in\Sigma_1$, and
$a\in\Sigma_0$.
Trees in $\T_\Sigma(X_J)$ are also called \emph{patterns}.
Of particular importance is the set of \emph{unary} patterns $\T_\Sigma(\{x_1\}) = \T_\Sigma(x_1)$.
This set forms a \emph{free monoid} where the monoid operation ``$\cdot$''
is substitution into the variable~$x_1$.
The tree $t= f(h(x_1), g(a,h(x_1)))$, e.g., can be uniquely factorized into $f(x_1,g(a,x_1))$ and $h(x_1)$.
We thus write
$f(h(x_1), g(a,h(x_1))) = f(x_1,g(a,x_1)) \cdot h(x_1)$.
We also consider the set $\C_\Sigma\subseteq\T_\Sigma(x_1)$ of \emph{contexts} over $\Sigma$
which is the subset of unary patterns which contain exactly one occurrence of $x_1$; 
technically, this means that each such context $t$ either is equal to $x_1$,
or is of the form $t=f(t_1,\ldots,t_k)$ for some $f\in\Sigma_k$, $k\geq 1$, and
$1\leq j\leq k$ so that $t_j$ is a context and $t_{j'}\in\T_\Sigma$ for all $j'\in\{1,\ldots,k\}\setminus\{j\}$.

In the following, $\Sigma$ and $\Delta$ denote fixed non-empty ranked alphabets of 
input and output symbols, respectively.
In this paper, we consider deterministic top-down tree transducers that are \emph{uniform-copying},
or \emph{uc}-transducers for short.
Intuitively, \emph{uniform-copying} means that each subtree of the input is processed at most once ---
while the produced output may be copied arbitrarily often. 
This restriction is trivially met by \emph{linear} top-down transducers -- but also by those that
arise from the top-down simulation of deterministic \emph{bottom-up} transducers 
by means of regular look-ahead.
Here, we refrain from introducing transducers with \emph{look-ahead} and \emph{inspection}
separately, as this would result in awkward duplication of almost identical definitions.
Instead, we find it convenient to introduce yet another model, namely, 
deterministic transducers with 
\emph{advice} --- which later can
be instantiated either with top-down deterministic inspection 
(no interference with the computation of the 
transducer, only restriction to relevant input) or bottom-up deterministic look-ahead
(interference with the computation as well as restriction to relevant input).

A \emph{finite tree automaton over $\Sigma$}, (for short, a \emph{TA}) $B$ 
is a tuple $(H,F,\delta)$ where
\begin{enumerate}
\item $H$ is a finite set of states,
\item $F\subseteq H$ is the set of accepting states, and
\item 
$\delta\subseteq\bigcup_{k\geq 0} H\times\Sigma_k\times H^k$
is the transition relation.
\end{enumerate}
%
%
The computation of $B$ on some input tree $t$ can be represented by 
a tree in $\T_T$ where the ranked alphabet $T$ consists of all 
transitions $\tau=\trans{h,f,h_1\ldots h_k}\in\delta$ where the rank of $\tau$
equals the rank of the input symbol $f$.
For $h\in H$, 
an $h$-computation $\phi$ for some $t=f(t_1,\ldots,t_k)\in\T_\Sigma$ is a tree 
$\phi=\tau(\phi_1,\ldots,\phi_k)$ where $\phi_i$ is an $h_i$-computation for $t_i$ for all
$i=1,\ldots,k$.
We write $h:t$  to indicate that there is an $h$-computation for $t$.
We write $\dom_B(h)=\{t\in\T_\Sigma\mid h:t\}$ and define
%
%
the set of trees accepted by $B$ as
$\L(B) = \{t\in\T_\Sigma\mid\exists h_0\in F\text{ such that }h_0:t\}.$
%
An $(h,h')$-computation $\phi$ of $B$ on some context $t\in\C_\Sigma$ is analogously defined 
as a context in $\C_T$ where $h$ is the state at the root and $h'$ is assumed at the variable
leaf.
We write $(h,h'): t$ to indicate that a $(h,h')$-computation for $t$ exists.
In particular, $(h,h): x_1$ for every state $h$ of $B$.
We assume that every occurring TA $B$ is \emph{trim}, i.e., it accepts a non-empty language
and every transition of $B$
occurs in some \emph{accepting} computation, i.e., some $h$-computation with $h\in F$.
We call $B$
\begin{itemize}
\item	\emph{bottom-up deterministic}, when for every tuple $(f,h_1\ldots h_k)\in\Sigma_k\times H^k$,
	there is at most one $h\in H$ so that $\trans{h,f,h_1\ldots h_k}\in\delta$;
\item	\emph{top-down deterministic} when $F$ consists of a single state only, and
	for every $k\geq 0$ and pair $(h,f)\in H\times\Sigma_k$,
	there is at most one tuple $(h_1\ldots h_k)\in H^k$ such that $\trans{h,f,h_1\ldots h_k}\in\delta$;
\item	\emph{unambiguous}, if for each $t\in\T_\Sigma$, there is at most one $h\in F$ and at most one
	$h$-computation $\phi$ of $B$ for $t$.
\end{itemize}
It is well-known that $B$ is unambiguous whenever
$B$ is bottom-up deterministic, or top-down deterministic.
%

A \emph{deterministic uniform-copying top-down tree transducer with advice
over $\Sigma$ and $\Delta$}
(for short, a \DBTB \emph{transducer}, or \emph{uc-transducer}, or a \DBTB) $A$ is a
tuple $(B,Q,\iota,\alpha,R)$
where
\begin{itemize}
\item	$B=(H,F,\delta)$
is an unambiguous \emph{advice} TA,
\item	$Q$ is a finite set of states together with a mapping $\iota:Q\to H$,
\item	$\alpha$ is a mapping which assigns to each final state $h\in F$ of $B$, an \emph{axiom}
	which is either a tree in $\T_\Delta$,
	or of the form $T_0 = p\cdot q_h(x_1)$ where $p\in\T_\Delta(x_1)$ with $q_h\in Q$
	and $\iota(q_h) = h$, and
\item	$R$ is the set of rules such that 
	for every transition $\trans{h,f,h_1\ldots h_k}\in\delta$ and every state 
	$q\in Q$ with $\iota(q) = h$, $R$ contains exactly one rule
	\begin{equation}
	q(f(x_1:h_1,\ldots,x_k:h_k))\to T	\label{eq:rule}
	\end{equation}
	where 
		$f\in\Sigma_k$ for some $k\geq 0$, and
		$T = p\,\{x_j\mapsto q_j(x_j)\mid j\in J\}$ where $p\in T_\Delta(X_J)$ for some subset
			$J\subseteq\{1,\ldots,k\}$ and for all $j\in J$,
		$q_j\in Q$ with $\iota(q_j) = h_j$
(thus $\{x_j\mapsto\dots\}$ is our notation of substituting leaves
labeled $x_j$ by the corresponding trees).
\end{itemize}
Intuitively, the transducer never gets stuck when the advice automaton can make a transition.
The \emph{size} $|A|$ of $A$ is defined as the sum of the sizes of its axioms, together with the
sum of the sizes of all rules of $A$ where the size of the rule \eqref{eq:rule} is $k+1$ plus the the size of the corresponding
right-hand side. Here, we allow trees occurring as axioms or right-hand sides to be represented as \emph{dags} 
in which common subtrees may be represented only once.
The size of such a dag is the sum of the numbers of its nodes and edges.
A \emph{uc}-transducer is \emph{linear}, if each input variable $x_i$ occurs
at most in the right-hand side of every rule (we also say ``\DLTB\ transducer'').

We remark that we view the set $Q$ of states of $A$ as symbols of rank 1 distinct from 
all symbols in $\Delta$.
%
%
Assume that $t = f(t_1,\ldots,t_k)\in\T_\Sigma$ with $h_j:t_j$ for $j=1,\ldots,k$.
Let $\tau = (h,f,h_1\ldots h_k)\in\delta$ be a transition
and $q\in Q$ a state with $\iota(q) = h$. 
Then we define the $q$-\emph{translation} 
$\sem{t}_{A,q}$
of $A$ for $t$ as follows.
Let 
\[
q(f(x_1:h_1,\ldots,x_k:h_k))\to p\{x_j\mapsto q_j(x_j)\mid j\in J\}
\]
denote the rule of $A$ for $q$ and $\tau$ where $J\subseteq\{1,\ldots,k\}$,
and $p\in\T_\Delta(X_J)$ is a pattern.
Then $\sem{t}_{A,q}$ is defined as
\[
\sem{t}_{A,q} = p\{x_j\mapsto \sem{t_j}_{A,q_j}\mid j\in J\}.
\]
Now additionally assume that $h\in F$ is an accepting state of $A$.
Then the \emph{translation} $\sem{t}_A$ of $A$ for $t$ is defined as follows.
\begin{itemize}
\item 	If $\alpha(h)\in\T_\Delta$ is ground, then $\sem{t}_A = \alpha(h)$;
\item	If $\alpha(h)= u\cdot q(x_1)$ for some $u\in\T_\Delta(x_1)$,
	then $\sem{t}_A = u\cdot\sem{t}_{A,q}$.
\end{itemize}
The \emph{translation} of $A$ thus is defined as the set
of pairs of trees:
\[
\{(t,\sem{t}_A)\mid t\in\T_\Sigma\text{ and }\exists h\in F\text{ such that }h:t \}.
\]
Technically, we also require the translations $\sem{t}_{A,q}^{h}$ 
and $\sem{t}_{A}^{h}$ provided by $A$ for a
context $t$ where $h$ is the state of the advice automaton $B$ 
for the single leaf node $x_1$ of $t$.
First, assume that $t = x_1$ is the empty context. 
Then $\sem{t}_{A,q}^{h} = q(x_1)$ whenever $\iota(q) = {h}$.
If additionally, ${h}\in F$ then $\sem{t}_A^{h} = \alpha({h})$ 
equals the axiom of $A$ for ${h}$.

Now assume that $t = f(t_1,\ldots,t_k)$ where $t_j\in\C_\Sigma$ is a context 
with $(h_j,h):t_j$
and $t_{j'}\in\T_\Sigma$ for $j'\in\{1,\ldots,k\}\setminus\{j\}$
with $h_{j'}:t_{j'}$.
Let again
\[
q(f(x_1:h_1,\ldots,x_k:h_k))\to p\{x_j\mapsto q_j(x_j)\mid j\in J\}
\]
denote the rule of $A$ for $q$ and $\tau$ where $J\subseteq\{1,\ldots,k\}$,
and $p\in\T_\Delta(X_J)$ is a pattern.
Then $\sem{t}_{A,q}^h$ is defined as
\[
\sem{t}_{A,q}^{h'} = p\{x_j\mapsto \sem{t_j}_{A,q_j}^{h'}, x_{j'}\mapsto\sem{t_{j'}}_{A,q_j}\mid j'\in J\setminus\{j\}\}.
\]
Now assume that $h'\in F$. 
If $\alpha(h')$ is ground, then $\sem{t}_{A}^{h'} = \alpha(h')$, and 
if $\alpha(h') = u\cdot q(x_1)$, then $\sem{t}_{A}^{h'} = u\cdot\sem{t}_{A,q}^{h'}$.
\ignore{
Given an $h$-computation $\phi$ of $B$ on some input tree $t\in\T_\Sigma$, 
the rule (in $R$) at each node of $t$ is uniquely determined by the state $q\in Q$ (with $\iota(q) = h$)
at the root of $t$, and the transitions chosen in $\phi$.
%
%
Assume that $t=f(t_1,\ldots,t_k)$, and 
$\phi=\tau(\phi_1,\ldots,\phi_k)$ for
$\tau=\trans{h,f,h_1\ldots h_k}$.
A $q$-computation $\psi$ of $A$ on $t$ with output $s$ 
is given by $\rho(\psi_1,\ldots,\psi_k)$ provided the following holds:
\begin{itemize}
\item	$\rho$ is a rule of the form \eqref{eq:rule};
\item 	if $j\in J$, then
	$\psi_j$ is a $q_j$-computation of $A$ on $t_j$ with some output $s_j$
	and otherwise, i.e., if $j\in\{1,\ldots,k\}\setminus J$, $\psi_j=\phi_j$;
\item  $s=T\{x_j\mapsto s_j\mid j\in J\}$.
\end{itemize}
If such a $q$-computation for $t$ exists with output $s$, we write $q:t\to s$.

As for TAs, we not only require the notion of a $q$-computation of $A$ on input trees 
$t\in\T_\Sigma$ with output $s$, but also the notion of a 
$(q,h)$-computation of $A$ on a context $t\in\C_\Sigma$ with
output $s$. Let $\phi$ denote a $(\iota(q),h)$-computation of $B$.
If $t=x_1$, then $x_1$ is a $(q,h)$-computation for $x_1$ with output $s= q(x_1)$ whenever
$\iota(q) = h$.
Now assume that $t= f(t_1,\ldots, t_k)$ and $t_j\in\C_\Sigma$ is a context,
and let $\phi=\tau(\phi_1,\ldots,\phi_k)$ denote the corresponding $(\iota(q),h)$-computation of $B$.
Assume that the rule $\tau$ is of the form \eqref{eq:rule}. 
Then $\tau(\psi_1,\ldots,\psi_k)$ is a $(q,h)$-computation for $t$ with output 
$s = p\{x_{j'}\mapsto s_{j'}\mid j'\in J\}$, 
if the following holds:
\begin{itemize}
\item 	If $j'\in\{1,\ldots,k\}\setminus J$, then $\psi_{j'} = \phi_{j'}$;
\item	If $j'\in J\setminus\{j\}$, then $\psi_{j'}$ is a $q_j$-computation for $t_{j'}$ with 
	output $s_{j'}$;
\item	If $j'=j\in J$, then
	$\psi_{j'}$ is a $(q_j,h)$-computation for $t_j$ with output $s_j$. 
\end{itemize}
We remark that if $s$ is non-ground, i.e., is not contained in $\T_\Delta$, 
then $s = s'\cdot q'(x_1)$ with $\iota(q') = h$.
If such a $(q,h)$-computation exists, we  write $(q,h):t\to s$.

The \emph{translation} of $A$ is the partial mapping $\sem{.}_A:\T_\Sigma\to\T_\Delta$ defined for every tree
with $h:t$ for some accepting state $h\in F$ where
$\sem{t}_A = p_0$ if the axiom of $A$ for $h$ is the ground tree $p_0$, and
$\sem{t}_A = p_0\cdot s$ if the axiom for $h$ equals $p_0\cdot q_0(x_1)$ and $q_0:t\to s$ holds.
Likewise, we define \emph{context translation} of $A$ for state $h_0\in F, h\in H$, as the partial mapping
$\sem{.}_A^{(h_0,h)}:\C_\Sigma\to(\T_\Delta\cup\T_\Delta(x_1)$ defined for every context $t$ with $(h_0,h):t$ by
$\sem{t}_A^{(h_0,h)} = p_0$ if the axiom of $A$ for $h_0$ is the ground tree $p_0$, and
$\sem{t}_A^{(h_0,h)} = p_0\cdot s$ if the axiom for $h_0$ equals $p_0\cdot q_0(x_1)$ and $(q_0,h):t\to s$ holds.
}
Let us briefly list instances of \emph{uc}-transducers with advice that are of interest here.
%

\textbf{Transducers with Look-ahead.}\quad
If the advice automaton $B$ is chosen as bottom-up deterministic, 
the transducer $A$ is a \emph{uc}-transducer with regular look-ahead
(for short, a \DBTR transducer, or a \DBTR).
%
%
%
%
%

\textbf{Transducers with Inspection}.\quad
If the advice automaton $B$ is chosen as top-down deterministic,
the transducer $A$ can be considered as a \emph{uc}-transducer with inspection
automaton $B$ (for short, a \DBTI\ transducer, or a \DBTI). 
In this case, the state annotations $h_i$ in the rule \eqref{eq:rule}, can be dropped since
these are obtained from $\iota(q)$ ($q$ the current state of the transducer) and the
input symbol $f$. 
%
A deterministic \emph{bottom-up} tree transducer in the classical sense, e.g., as 
in~\cite{DBLP:journals/mst/Engelfriet75} is obtained in our model as a \DBTB\ 
transducer where $Q=H$ and $\iota$ is the identity.
A classical deterministic \emph{top-down} tree transducer with uniform-copying, 
on the other hand, is obtained as a \DBTI\ transducer where 
the inspection does not restrict the domain. This can be achieved, e.g., by setting
$H= Q\cup\{\top\}$ for a fresh symbol $\top$ and $\iota(q) = q$.
Moreover, for each $f\in\Sigma_k$, $\trans{\top,f,\top^k}\in\delta$ as well as
$\trans{q,f,q'_1\ldots q'_k}\in\delta$
whenever there is a rule $q(f(x_1,\ldots,x_k)) \to T$ such that
for each $i=1,\ldots,k$, $q'_i= q_i$ if $q_i(x_i)$ occurs in $T$ and $q'_i = \top$ otherwise.
Alternatively, in that case the advice automaton can be removed entirely from the definition of the transducer -- thereby
obtaining the classical model of deterministic top-down transducer (as defined, e.g., in \cite{DBLP:journals/mst/Rounds70}), 
here with the \emph{uniform-copying} restriction.
	
We remark that in the same way, deterministic \emph{linear} top-down 
tree transducers with look-ahead as well as deterministic linear bottom-up transducers and
deterministic linear top-down transducers with inspection are instances of \emph{linear} \DBTB transducers.

\ignore{
E.g., if $\tau\;\equiv\;q(f(x_1:h_1,x_2:h_2)) \to h(q(x_1))$, then
	$\tau(\phi,\psi)$ is a computation where $\phi$ is an execution starting
in state $q_1$ and $\psi$ is a computation of the look-ahead automaton resulting in state $h_2$. 
If $\phi$ is a $q_1$-computation for $t_1$ with output $s_1$, and
$\psi$ a $h_2$-computation for $t_2$, then
$\tau(\phi,\psi)$ is a $q$-computation for $t=f(t_1,t_2)$ with output $s=h(s_1)$.
In general, we write
\[
\tau(\phi,\psi)\vdash_q t\to s
\]
Now, an \emph{accepting} computation $\phi$ for $t$ with output $s$ starts in some axiom $p$. 
If $p$ is ground and corresponds to look-ahead state $h$, then
$\phi= p\cdot\psi'$ 
if $s=p$ and $\psi'\vdash_h t$.
If $p = p'\cdot q_0(x_1)$, then $\phi=p\cdot\phi'$ and $s=p'\cdot s'$, if $\phi'$ is the
$q_0$-computation $\phi'$ on $t$ and 
$\phi\vdash_{q_0} t\to s'$. 
In these cases, we write $\phi\vdash_A t\to s$.
%

IDEA: replace single variable in input tree with state $h$ of the look-ahead automaton!

%
%
This notion extends to \emph{contexts} $t\in\C_\Sigma$ for some look-ahead state $h'$
for the variable leaf in $t$. In this case, we write
\[
\phi\vdash_{q,h'} t\to s
\]
where now $s$ may either be ground, or contain calls $q'(x_1)$ for some $q'$ with
$\iota(q') = h'$.
For look-ahead automata, we instead write $\psi\vdash_{h,h'} t$, if $\psi$ is an $h$-computation
for $t$ with $h'$ at $x_1$.
}

\section{A Dedicated Earliest Normal Form for \emph{uc}-Transducers with Advice}
	\label{s:earliest}

In this section we present constructions of earliest normal-forms for 
\emph{uc}-transducers and \emph{linear} transducers with advice.
In the following, we fix some trim unambiguous TA $B$ for advice.
A construction of \emph{earliest} deterministic top-down transducers has already been provided 
in~\cite{DBLP:journals/tcs/EngelfrietMS16} for top-down deterministic domain automata $B$ and as well as 
in~\cite{DBLP:conf/icgi/BoiretLN16} for bottom-up deterministic $B$. 
Here, we are slightly more liberal by allowing \emph{unambiguous} $B$ to generalize both cases.
The constructions from~\cite{DBLP:journals/tcs/EngelfrietMS16,DBLP:conf/icgi/BoiretLN16}, on the other hand, 
neither preserve linearity nor uniform-copying.
\begin{example}\label{e:one}
Consider a linear top-down transducer with the rules:
\[
{
\begin{array}{lcllcl}
q_0(g(x_1)) 	&\to& q_1(x_1)&
q_0(a) &\to& a	\\
q_1(f(x_1,x_2))	&\to& f(q_0(x_1), q_0(x_2))\qquad\qquad&
q_0(b) &\to& b
\end{array}}
\]
and the axiom $q_0(x_1)$. 
%
The (canonical) earliest transducer constructed according to the methods 
in~\cite{DBLP:journals/tcs/EngelfrietMS16} has the same axiom $q_0(x_1)$, but the rules:
\[
{
\begin{array}{lcllcl}
q_0(g(x_1)) 	&\to& f(q_{11}(x_1), q_{12}(x_1))\ \ \qquad&
q_0(a) &\to& a	\\
q_{11}(f(x_1,x_2))	&\to& q_0(x_1)&
q_0(b) &\to& b	\\
q_{12}(f(x_1,x_2))	&\to& q_0(x_2)
\end{array}}
\]
where no inspection automaton is required.
The non-linearity in the first rule 
arises inevitably, because the output $f$-node is already determined at this point
and therefore must be output for the transducer to be \emph{earliest}.
Since the transductions corresponding to states $q_{11}$ and $q_{12}$ differ, 
the resulting
transducer is not only non-linear, but also not \emph{uniform-copying}.
\qed
\end{example}
We introduce \emph{dedicated} constructions which
allow to construct equivalent \emph{canonical} transducers, but retain linearity or
uniform-copying.
\ignore{
We introduce a \emph{dedicated} construction which
allows constructing equivalent \emph{canonical} transducers, but retains both linearity and
uniform-copying.
}

It turns out that these normal forms can be obtained in polynomial time.
Our key insight is that \emph{dedicated} notions should be provided 
for the notion of \emph{maximal common prefix} of a given set $S$ of trees. 
In~\cite{DBLP:journals/tcs/EngelfrietMS16}, a pattern 
such as $p = a(x_1,g(x_2))$ was used to represent the common part of trees in $S$
(note that they do not use distinct variables, but one symbol
$\top$ to denote \emph{any tree} and whose distinct occurrences can be substituted independently).
This meant for an element $t\in S$ such as $t= a(b,g(c))$ that 
the variables in $p$ could correspond to not necessarily isomorphic
subtrees of $t$, in the example, $b$ and $c$, respectively. 
In that point, we will now be more restrictive and only allow a single variable $x_1$ and \emph{substitutions},
i.e., equal replacements of all occurrences of $x_1$ in patterns.
To distinguish these patterns from those in~\cite{DBLP:journals/tcs/EngelfrietMS16}, 
we call them \emph{uniform}.
Let $\Delta$ be a ranked alphabet.
Let us denote by $\P_{\Delta}$
the set $\T_\Delta\cup\T_\Delta(x_1)\cup\{\bot\}$ of all ground trees 
and unary patterns, extended with one specific fresh element $\bot$.
This set forms a \emph{partial order} 
where for $t_1,t_2\in\P_\Delta$,
$t_1\sqsubseteq t_2$ iff  $t_1=\bot$ or $t_1 = t_2\{x_1\mapsto s\}$ for 
some $s\in\T_\Delta\cup\T_\Delta(x_1)$.
In fact, $\P_{\Delta}$, partially ordered in this way, 
forms a \emph{complete lattice} with finite ascending chains. 
In particular, the top-most element is $x_1$, and the binary least upper
bound operation $\sqcup$ for incomparable elements $t_1,t_2\neq\bot$, is given by $t_1\sqcup t_2 = s$ 
where $s$ is the maximal tree in $\T_\Delta(x_1)$ such that
$s\{x_1\mapsto t'_i\} = t_i$ for suitable trees $t'_i$ ($i=1,2$).
%

We remark that uniform patterns may contain more than one occurrence of $x_1$ --- 
all representing, though, isomorphic subtrees.
Let $\P_\Delta^{(1)}\subseteq\P_\Delta$ denote the subset $\T_\Delta\cup\C_\Delta\cup\{\bot\}$
of all elements which either equal $\bot$ or contain at most \emph{one} occurrence of $x_1$.
Patterns in that set are also called \emph{1-patterns}.
%
%
For the induced partial ordering on $\P_\Delta^{(1)}$, 
we again obtain a complete lattice
with finite ascending chains. For a distinction, let us denote the least upper bound 
operation with respect to $\P_\Delta^{(1)}$ with $\sqcup^{(1)}$.

\begin{example}\label{ex:cl}
The difference between the two least upper bound operations becomes 
apparent when considering trees which differ in more than one subtree:
\[
\begin{array}{lll}
f(g(a,a),c)\sqcup f(g(b,b),c) &=&	f(g(x_1,x_1),c)	\\
f(g(a,a),c)\sqcup^{(1)} f(g(b,b),c) &=&	f(x_1,c).	
\end{array}
\]
On the other hand, 
$f(g(a,a),c)\sqcup f(g(b,b),d)=f(g(a,a),c)\sqcup^{(1)} f(g(b,b),d)=x_1$.	
We remark that the earliest construction in~\cite{DBLP:journals/tcs/EngelfrietMS16}
would return $f(g(\bot,\bot),\bot)$ for the latter example --- implying
that the place holder $\bot$ no longer represents \emph{isomorphic} subtrees.
\qed
\end{example}

\noindent
%

\subsection{Computing Maximal Common Prefixes}
\label{ss:pref}
For the rest of this section, let $A=(B,Q,\iota,R,\alpha)$ denote a fixed \emph{uc}-transducer 
with advice.
For $q\in Q$, let 
\[
\begin{array}{lll}
\pref_A(q) &=& \bigsqcup\{\sem{t}_{A,q}\mid t\in\dom_B(\iota(q)\}	\qquad
 					\text{ and }	\\
\pref^{\,(1)}_A(q) &=& \bigsqcup^{(1)}\{\sem{t}_{A,q}\mid t\in\dom_B(\iota(q))\}.
\end{array}
\]
We remark that even for a \emph{linear} transducer, the values $\pref_A(q)$ and $\pref_A^{\,(1)}(q)$ may indeed differ.

\begin{example}
Let $q$ be a state of a linear transducer $A$ with the following two rules:
\[
\begin{array}{lcl}
q(f(x_1:h_1)) &\to& g(a,a)\\
q(f(x_1:h_2)) &\to& g(b,b).
\end{array}
\]
Then $\pref_A(q)=g(x_1,x_1)$ while $\pref_A^{\,(1)}(q)=x_1$.
This means that a corresponding earliest \emph{uc}-transducer outputs
a $g$-node in state $q$, while a corresponding earliest \emph{linear} transducer
produces no output node in state $q$.
\qed
\end{example}

In the following, we show that  $\pref_A: Q\to\P_\Delta$ 
as well as $\pref_A^{\,(1)}:Q\to \P^{(1)}_\Delta$ (in case that $A$ is in fact linear)
can be obtained 
as the least solutions of the sets of constraints $\C_A$ and $\C_A^{(1)}$, respectively.
%

The sets $\C_A$, $\C^{(1)}_A$ consist of one constraint $c(\tau)$, $c^{(1)}(\tau)$ for each rule $\tau$
of $A$. 
Assume that $\tau$ is the rule
$q(f(\ldots)) \to T$ of $A$ where $T = p\{x_j\mapsto q_j(x_j)\mid j\in J\}$ for some 
$J\subseteq{\mathbb N}, p\in\T_\Delta(X_J)$, and suitable $q_j\in Q$. Then the constraint $c(\tau)$ is given by
\begin{eqnarray}
\sigma(q)&\sqsupseteq& \sem{T}\sigma\qquad \label{eq:rule_sharp}	
\end{eqnarray}
while the constraint $c^{(1)}(\tau)$ is given by
\begin{eqnarray}
\sigma^{(1)}(q)&\sqsupseteq& \sem{T}^{(1)}\sigma^{(1)}.\qquad \label{eq:rule_sharp1}	
\end{eqnarray}
Here, $\sigma$ (and $\sigma^{(1)}$) are mappings from states in $Q$
to patterns in $\T_\Delta(x_1)$.
The right-hand side $\sem{T}\sigma$ returns $\bot$ if for any state $q_j,j\in J$, 
$\sigma(q_j) = \bot$. Otherwise,  
assume that $p'$ is obtained from $p$ and $\sigma$ by replacing $q_j(x_j)$
with $\sigma(q_j)\cdot x_j$
where $j\in J$ and with $\sigma(q_j)$ whenever $\sigma(q_j)$ is ground.
If $p'$ is ground, we set $\sem{T}\sigma = p'$.
If $p'$ contains occurrences of $x_j$ for a single $j\in J$,
i.e., is of the form $p' = p''\cdot x_j$ for some $p''\in\T_\Delta(x_1)$,
then $\sem{T}\sigma = p''$.
Otherwise, i.e., if $p'$ contains occurrences of more than one variable,
then $\sem{T}\sigma = u$ where $u\in\T_\Delta(x_1)$ is the maximal prefix
such that $p' = u\cdot p''$ for some $p''$ containing all $x_j$, i.e., $u$ has maximal size with
	this property and thus is \emph{least} with respect to the ordering on patterns.

Likewise for linear transducers, the right-hand side $\sem{T}^{(1)}\sigma^{(1)}$ is evaluated 
in the same way -- only that now prefixes are determined w.r.t.\ the patterns in $\P_\Delta^{(1)}$.

\begin{example}
Let $T = f(g(q_2(x_2),a), g(q_2(x_2), q_1(x_1))$, and thus
$p= f(g(x_2,$ $a), g(x_2,x_1))$.
Assume that we are given $\sigma = \{q_1\mapsto a, q_2\mapsto b(x_1)\}$.
Then we have that 
\[
p'=f(g(b(x_2),a), g(b(x_2),a))
\]
and thus,
\[
\sem{T}\sigma = f(x_1,x_1)\cdot g(x_1,a)\cdot b(x_1) 
	=f(g(b(x_1),a),g(b(x_1),a)).
	\]
\qed
\end{example}
Each right-hand side of a constraint in $\C_A$ ($\C^{(1)}_A$)
represents a function which is \emph{distributive} in each argument, 
i.e., commutes with the binary least upper bound operator in each accessed argument $\sigma(q_j)$ ($\sigma^{(1)}(q_j)$).
Recall that any distributive function is also monotonic.
Since the partial orderings on $\P_\Delta$ 
and likewise on $\P_\Delta^{(1)}$ 
are complete lattices
with finite ascending chains,
the constraint systems \eqref{eq:rule_sharp} and \eqref{eq:rule_sharp1}
have a least solution. 
We thus obtain:

\begin{lemma}\label{l:patterns}
Let $\pref_A(q),q\in Q$, and $\pref_A^{\,(1)}(q), q\in Q)$, denote the least solutions of the set of constraints 
\eqref{eq:rule_sharp} over  the complete lattice $\P_\Delta$ and 
\eqref{eq:rule_sharp1} over  the complete lattice $\P_\Delta^{(1)}$ (in the case that $A$ is linear), respectively. Then
for every $q\in Q$ with $\iota(q) = h$,
\begin{eqnarray}
\pref_A(q) &=& \bigsqcup\{\sem{t}_{A,q}\mid t\in\dom_B(h)\}
\label{eq:pref}	\\
\pref_A^{\,(1)}(q) &=& \bigsqcup^{(1)}\{\sem{t}_{A,q}\mid t\in\dom_B(h)\}
\label{eq:pref1}
\end{eqnarray}
holds.
Moreover, the least solution can be computed in polynomial time.	
\end{lemma}
\begin{proof}
Recall that by our assumption, the advice automaton is trim. Therefore, 
$\dom_B(\iota(q))$ is non-empty, i.e., $\pref_A(q)\neq\bot$ for each $q\in Q$.
The equality in equation~\eqref{eq:pref} (and in equation~\eqref{eq:pref1})
then is due to the Fixpoint Transfer Lemma
\cite{DBLP:journals/jacm/AptP86}.
More explicitly,
let $X_q^{(i)}$ denote the $i$th iterate of the fixpoint iteration for the constraint
system \eqref{eq:rule_sharp} for $i\geq 0$. 
By induction on $i$, it can be verified that $X_q^{(i)}$ equals the maximal
common prefix of all $\sem{t}_{A,q}$ for trees $t\in\dom_B(\iota(q))$ of depth less than $i$.
In case $A$ is linear, we may define $X_q^{(i)}$ analogously as the $i$th iterate of the fixpoint iteration for the constraint system
\eqref{eq:rule_sharp1}.

Thereby, the prefixes $X_q^{(i+1)}$ can be determined from the prefixes $X_q^{(i)}$
in polynomial time. This is obvious for \emph{linear} transducers $A$ and the constraint system
\eqref{eq:rule_sharp1}.
When $A$ is \emph{uniform-copying}, and general uniform patterns are used, 
polynomial time can be obtained when 
trees are represented as \emph{dags} where isomorphic subtrees are represented only once.
Concerning the complexity of the construction.
We claim that 
the number of iterations required for reaching the least fixpoint of the constraint system
is bounded by a polynomial in the size $|A|$ of the transducer $A$.
%
%
A first value different from $\bot$ occurs after at most $n$ iterations where $n$ is the number of states of $A$.
This value can be represented by some dag composed of right-hand sides where no right-hand side occurs more than once. 
In each further iteration, the value for at least one state $q$ shrinks (or the algorithm terminates). 
In total, the least fixpoint is reached after at most $n\cdot|A|$ iterations. Since the representation of each intermediate 
tree is bounded by $|A|$, the overall complexity statement follows.
\end{proof}

A similar constraint system is used for the construction of maximal common prefixes in \cite{DBLP:journals/jcss/EngelfrietMS09} for unrestricted deterministic top-down tree transducers.
Due to the more general form of patterns, however, the construction there cannot rely on compressed representations of occurring
trees -- implying that no polynomial upper time bound has been provided.

\subsection{Earliest Transducers}\label{ss:earliest}

Let $A$ denote some \emph{uc}-transducer (\emph{linear} transducer) $A$ with advice.
%
%
We call $A$ an earliest \emph{uc}-transducer (\emph{linear} transducer), if
$\pref_A(q) = x_1$ 
($\pref_A^{(1)}(q) = x_1$)
for all states $q$ of $A$.
If this is not yet the case, 
we construct a transducer $A'$ of the same kind as $A$ as follows
where we only present the construction for \emph{uc} transducers (the linear case is analogously obtained where 
each $\pref_A$ is replaced with $\pref^{\,(1)}_A$).
The set $Q'$ of states of $A'$ consists of the \emph{non-constant} states of $A$, i.e., of all 
states of $A$ with at least two distinct output trees. This set is given by
$Q' = \{q\in Q\mid \pref_A(q)\not\in\T_\Delta\}$.
%
The mapping $\alpha'$ of axioms of $A'$ is obtained from the mapping $\alpha$ of axioms of $A$ 
as follows. Let $h$ denote a final state of $B$. If $\alpha(h)$ is ground, then $\alpha'(h) = \alpha(h)$.
Otherwise, $\alpha(h) = p\cdot q(x_1)$ for some $p\in\T_\Delta(x_1)$ and $q\in Q$.
Assume that $\pref_A(q) = s$. If $s\in\T_\Delta$, then $\alpha'(h) = p\cdot s$; and if
$s\in\T_\Delta(x_1)$, then $\alpha'(h) = p\cdot s\cdot q(x_1)$. 

Now assume that $q\in Q'$, and $\pref_A(q)=u$.
Then for each rule $q(f(x_1:h_1,\ldots,x_k:h_k))\to p\{x_j\mapsto q_j(x_j)\mid j\in J\}$ of $A$,
$A'$ has a rule $q(f(x_1:h_1,\ldots,x_k:h_k))\to T'$ where $T'$ is defined as follows.
For $j\in J$, let $s_j=\pref_A(q_j)$ if $\pref_A(q_j)\in\T_\Delta$,
and $s_j= u_j\cdot q_j(x_j)$ if $\pref_A(q_j) = u_j\in\T_\Delta(x_1)$.
Then $u$ must be a prefix of 
$p\{x_j\mapsto s_j\mid j\in J\}$, and we choose $T'$ such that
$p\{x_j\mapsto s_j\mid j\in J\} = u\cdot T'$ holds.

\begin{lemma}\label{l:earliest}
Assume that $A$ is a \DBTB\ (\DLTB) 
without constant states
and $A'$ the \DBTB\ 
(\DLTB) 
as constructed above. Then,
\begin{enumerate}
\item	For each state $q$ of $A'$ with $u=\pref_A(q)$ ($u=\pref_A^{\,(1)}(q)$) with $\iota(q) = h$, it holds that
	\begin{enumerate}
	\item	$\sem{t}_{A,q} =u\cdot \sem{t}_{A',q}$ for all $t\in\dom_B(h)$;
	\item	$\pref_{A'}(q) = x_1$ ($\pref^{\,(1)}_{A'}(q) = x_1$) holds for all states $q$
		of $A'$.
	\end{enumerate}
\item	$A'$ is equivalent to $A$.
\item	$A'$ can be constructed from $A$ in polynomial time.
\end{enumerate}
\end{lemma}

\begin{proof}
Properties 1.a and 1.b of Lemma \ref{l:earliest} can be proven by structural 
induction on the input tree $t$.
Property 2 then follows from 1.a and 1.b by the definition of the axioms of $A'$.
For property 3 we allow ourselves that both $A$ and $A'$ use right-hand sides represented by means of 
\emph{ordered dags} which represent 
identical subtrees only once.
According to Lemma \ref{l:patterns}, the maximal prefixes $\pref_A(q)$ ($\pref_A^{\,(1)}(q)$) can be
computed in polynomial time. Since the required factorizations of right-hand sides can be computed on 
compressed representations in polynomial time, also property 4 follows.
\end{proof}

\noindent
The corresponding construction in \cite{DBLP:journals/jcss/EngelfrietMS09} and its proof are 
more complicated, because each state of the original transducer may be split into multiple new states 
of the corresponding earliest transducer (see Example \ref{e:one}).

%
The earliest transducer constructed for Lemma \ref{l:earliest} may further be \emph{minimized}.
For that, we introduce $\equiv$ as the \emph{coarsest} equivalence
relation  $\approx$ on states such that $q\approx q'$ implies that 
\begin{enumerate}
\item[\textbf{(M)}]	$\iota(q) = \iota(q')$, 
		and for each transition $(h,f,h_1\ldots h_k)$ 
		with $\iota(q) = h$ and rules
		$q(f(x_1:h_1,\ldots,x_k:h_k)) \to p\{x_j\mapsto q_j(x_j)\mid j\in J\}$,
		$q'(f(x_1:h_1,\ldots,x_k:h_k)) \to p'\{x_{j'}\mapsto q'_{j'}(x_{j'})\mid {j'}\in J'\}$
		of $A$ with patterns $p\in\T_\Delta(X_J)$, $p'\in\T_\Delta(X_{J'})$,
		it holds that $J=J'$, $p = p'$ and $q_j\approx q'_j$ for all $j \in J$.
\end{enumerate}

\begin{lemma}\label{l:canonical}
Let $A$ be an earliest \emph{uc}-transducer (\emph{linear} transducer) and $\equiv$ 
the equivalence relation as defined above.
Then the following holds:
\begin{enumerate}
\item	$q\equiv q'$ iff $\iota(q) =\iota(q')$, and $\sem{t}_{A,q} = \sem{t}_{A,q'}$ for all $t\in\dom_B(\iota(q))$;
\item	$\equiv$ can be constructed in polynomial time.
\end{enumerate}
\end{lemma}


\begin{proof}
The proof of Lemma~\ref{l:canonical} follows closely the corresponding 
proof of Theorem~13 of~\cite{DBLP:journals/tcs/EngelfrietMS16}.
%
We only prove the first statement.
Assume that we are given an equivalence relation $\approx$ which satisfies \textbf{(M)}.
By definition, $q\approx q'$ implies that also $\iota(q) =\iota(q')$ holds. 
We proceed by structural induction on the input tree $t = f(t_1,\ldots,t_k)$.
Assume that $\sem{t}_{A,q} = s$. This means that there is a 
transition $\trans{h,f,h_1\ldots h_k}$ of $B$ for $h=\iota(q)$ and a rule
\[
q(f(x_1:h_1,\ldots,x_k:h_k))\to p\{x_j\mapsto q_j(x_j)\mid j\in J\}
\]
for some pattern $p\in\T_\Delta(X_J)$ such that
$s = p\{x_j\mapsto\sem{t_j}_{A,q_j}\mid j\in J\}$.
Now assume that $q\approx q'$. Since $\iota(q') = h$ as well,
there is one rule
\[
q'(f(x_1:h_1,\ldots,x_k:h_k))\to p'\{x_j\mapsto q'_j(x_j)\mid j\in J'\}
\]
Because of property \textbf{(M)},
$J = J'$, $p = p'$ and $q_j\approx q'_j$ for all $j\in J$.
By inductive hypothesis therefore, $\sem{t_j}_{A,q_j} = \sem{t_j}_{A,q'_j}$ for all $j\in J$, and
we have that
\[
\sem{t}_{A,q} = s 
= p\{x_j\mapsto\sem{t_j}_{A,q_j}\mid j\in J\}
= p'\{x_j\mapsto\sem{t_j}_{A,q'_j}\mid j\in J'\}
= \sem{t}_{A,q}
\]
-- which we wanted to prove.
For the reverse	implication, we define an equivalence relation $\sim$ on the set $Q$ 
by $q\sim q'$ iff $\iota(q) = \iota(q')$ and for all $t\in\dom_B(\iota(q))$,
$\sem{t}_{A,q} = \sem{t}_{A,q'}$. We claim that the relation $\sim$ 
satisfies property \textbf{(M)}.
Since $\equiv$ was chosen as the \emph{coarsest} equivalence relation satisfying property \textbf{(M)},
the first claim of the lemma follows.
In order to prove that $\sim$ satisfies property \textbf{(M)},
consider a pair $q,q'$ with $q\sim q'$ with $\iota(q) = \iota(q') = h$,
and a transition $\trans{h,f,h_1\ldots h_k}$ with rules
\[
q(f(x_1:h_1,\ldots,x_k:h_k))\to p\{x_j\mapsto q_j(x_j)\mid j\in J\}
\]
and
\[
q'(f(x_1:h_1,\ldots,x_k:h_k))\to p'\{x_{j'}\mapsto q'_{j'}(x_{j'})\mid {j'}\in J'\}
\]
for patterns $p\in\T_\Delta(X_J)$, $p'\in\T_\Delta(X_{J'})$. Subsequently, we show that 
necessarily, $J=J',p=p'$, and $q_j\sim q'_j$ holds for all $j\in J$.
Consider an input tree $t$ of the form $t=f(t_1,\ldots,t_k)$
so that $t\in\dom_{B}(h)$ and thus also $t_j\in\dom_B(h_j)$ for all $j=1,\ldots,k$.
Let $\sem{t}_{A,q} = \sem{t}_{A,q'}\;{=:}\;s$.
We first prove that indeed $J=J'$ and $p=p'$ holds. 
Assume that this were \emph{not} the case. 
Since
\[
p\{x_j\mapsto\sem{t}_{A,q_j}\mid j\in J\} = p'\{x_{j'}\mapsto\sem{t}_{A,q'_{j'}}\mid j'\in J'\}.
\]
there is a maximal pattern $u\in\T_\Delta(X_I)$ for some set $I=\{1,\ldots,r\}$
with
\[
p = u\{x_i\mapsto p_i\mid i\in I\} \quad\text{and}\quad p' = u\{x_{i}\mapsto p'_i\mid i\in I\}
\]
so that $p_i,p'_i$ differ.
We claim that then for every $i$, the following holds:
\begin{enumerate}
\item	the sets of variables occurring in $p_i$ and $p'_i$, respectively, agree.
\item	either $p_i$ or $p'_i$ is a variable.
\end{enumerate}
Consider the first claim. Assume that $p_i\in\T_\Delta(J_i)$.
If $j\in J_i$, i.e., $x_j$ occurs in $p'_i$, this means that the output for trees in $\dom{B}(h_j)$
cannot be constant (otherwise, due to the earliest construction, 
no variable $x_h$ occurred anywhere in $p$). Let $t'_j\in\dom_B(h_j)$ a tree so that
$\sem{t'_j}_{A,q_j} \neq\sem{t_j}_{A,q_j}$. But then, if $x_j$ does not occur $p'_j$, 
the outputs for $f(t_1,\ldots,t_{j-1},t'_j,t_{j+1},\ldots,t_k)$ by $q$ and $q'$ must differ in the subtree
corresponding to the pattern variable $x_i$ in $u$.

For the second claim we note that $\sem{t}_{A,q}$ and $\sem{t}_{A,q'}$ necessarily differ if
$p_i,p'_i$ both have distinct output symbols at their roots. If on the other hand, 
$p_i = g(v_1,\ldots,v_m)$ and $p'_i = g(v'_1,\ldots,v'_m)$ for the same output symbol $g$
and some $v_j,v'_j$, then $u$ would not be the
\emph{maximal} common output pattern. Therefore, the second claim follows as well.

From the two claims, we conclude that one of the $p_i,p'_i$ must equal a variable $x_j$, while the other
is of the form $v\cdot x_j$ for some $v\neq x_1$ (which, in case that $A$ is an \DLTB\ is linear). 
W.l.o.g., let $p_i = x_j$.
Then $v$ is a common prefix of all outputs produced by $q_j$ for input trees from $\dom_B(h_j)$.
Since by our assumption, $A$ is earliest, $v$ must equal $x_1$, i.e., $p_i = p'_i$:
contradiction.

In summary, we have shown that $J=J'$ and $p=p'$ necessarily must hold.
But then for all $j\in J$,
$\sem{t'_j}_{A,q_j} =_\sem{t'_j}_{A,q'_j}$ also must hold for all $t_j\in\dom_B(h_j)$ --
implying that $q_j\sim q'_j$ for all $j\in J$.
Therefore, $\sim$ indeed satisfies property \textbf{(M)}.
This completes the proof.
\end{proof}

\noindent
Let $A = (B,Q,\iota,R,\alpha)$ denote a \DBTB\ (\DLTB) transducer, 
and $A'= (B,Q',\iota,$ $R',\alpha')$ 
and $A'' = (B,Q'',\iota'',R'',\alpha'')$ the corresponding
earliest and earliest canonical transducers. 
For each 
state $q\in Q'$ of $A'$, there is a suffix
$d_q\in\T_\Delta(x_1)$ such that the output produced by $A$ when reaching state $q$ lags behind the output already produced by $A'$
when reaching the same state $q$ by $d_q$. 
In fact, $d_q=\pref(q)$  ($d_q=\pref^{\,(1)}_A(q)$) is the prefix provided by the least solution of the constraint system
\eqref{eq:rule_sharp} (or \eqref{eq:rule_sharp1}) for transducer $A$ and state $q$.
Technically, this means for every context $c$ that 
$\sem{c}_{A}^h = u\cdot q(x_1)$ for some $u\in\T_\Delta(x_1)$ iff 
$\sem{c}_{A'}^h = u'\cdot q(x_1)$ for some $u'\in\T_\Delta(x_1)$ 
where $u\cdot d_q = u'$.

The transducer $A''$ then is obtained from $A'$ by collapsing multiple states into one. This means that there is a surjective 
mapping $\pi:Q'\to Q''$ so that $q$ in $A'$ and $\pi(q)$ in $A'$ are equivalent, and
for every context $c$ by which $A'$ reaches some state $q$, $A''$ reaches the state $\pi(q)$
while producing the same output. 
Putting these two mappings together, we define the \emph{aheadness} map for $A$
$\textsf{ah}$. Let $q\in Q$ with $\iota(q) = h$.
If $q$ is constant where $s = \sem{t}_{A,q}$ for all $t\in\dom_B(\iota(q))$, then we set $\textsf{ah}(q) = s$.
Otherwise, we set $\textsf{ah}(q) = u \cdot q''(x_1)$ if $u=\pref_A(q)$ and $q''=\pi(q)$.
%
Altogether from Lemmas~\ref{l:earliest} and~\ref{l:canonical}, we obtain:

\begin{theorem}\label{t:canonical}
For each \DBTB\ (\DLTB) transducer $A$, a unique minimal earliest
\DBTB\ (\DLTB) transducer $A'$ can be constructed such that 
\begin{enumerate}
\item
the aheadness map for $A$ can be constructed in polynomial time.	
\item	
$A'$ has at most as many states as $A$,
\item	
$A'$ is equivalent to $A$ where
\item	
$A'$ can be constructed in polynomial time.
\qed
\end{enumerate}
\end{theorem}

\noindent
As an immediate corollary we obtain:

\begin{corollary}\label{c:poly-eq}
Assume that the input alphabet is fixed.
Then equivalence for \DBTB\ as well as for \DLTB\ transducers is decidable in polynomial time.
\end{corollary}

\begin{proof}
Assume that we are given \DBTB\ transducers $A_1,A_2$ \DBTB\ transducers with unambiguous advice automata 
$B_1,B_2$, respectively. First, we check whether $\L(B_1) = \L(B_2)$.
This problem can be decided in polynomial time \cite{Seidl90}.
If $\L(B_1)\neq\L(B_2)$, then $A_1$ and $A_2$ cannot be equivalent.
So, assume that $\L(B_1) = \L(B_2)$. Let $B=B_1\times B_2$ denote the product automaton of $B_1,B_2$
where the set $F$ of accepting states of $B$ consists of all pairs $(f_1,f_2)$, $f_i$ accepting state of
$B_i$ so that $\dom_{B_1}(f_1)\cap\dom_{B_2}(f_2)\neq\emptyset$. 
The set $H$ of all states of $B$ is given by all pairs $(h_1,h_2)$ of states of $B_1$ and $B_2$
so that $\dom_{B_1}(h_1)\cap\dom_{B_2}(h_2)\neq\emptyset$ and $(h_1,h_2)$ is jointly reachable from
some $(f_1,f_2)\in F$, i.e., $F\subseteq Q$, and if $(h^{(1)},h^{(2)})\in Q$ and
$\trans{h^{(i)},f,h^{(i)}_1\ldots h^{(i)}_k}$ is a transition of $B_i$ for $i= 1,2$, with
$\dom_{B_1}(h^{(1)}_j)\cap \dom_{B_1}(h^{(2)}_j)\neq\emptyset$ for all $j=1,\ldots,k$,
then 
$\trans{(h^{(1)},h^{(2)}),f,(h^{(1)}_1,h^{(2)}_1)\ldots(h^{(1)}_k,h^{(2)}_k)}$ is a transition of 
$B$ with $(h^{(1)}_j,h^{(2)}_j)\in Q$ for all $j$.
Since both $B_1$ and $B_2$ are unambiguous, $B$ is unambiguous as well with 
$\L(B_i) = \L(B_1)\cap\L(B_2) = \L(B)$ for $i=1,2$.

For the transducers $A_1,A_2$, we construct transducers $A'_1,A'_2$ with advice $B$. The
set of states of $A'_1$ consists of pairs $\angl{q,h_2}$ where $q$ is a state of $A_1$ and 
$(\iota(q),h_2)$ is a state of $B$. Then we set $\iota'_1(\angl{q,h_2}) = (\iota_1(q),h_2)$.
\begin{itemize}
\item	If $\alpha_1$ provides the axioms for $A_1$, then $A'_1$ has as axiom  the mapping
	$\alpha'_1$ where
	$\alpha'_1(f_1,f_2) = \alpha_1(f_1)$ if $\alpha_1(f_1)\in\T_\Delta$, and
	$\alpha'_1(f_1,f_2) = p\cdot\angl{q,f_2}(x_1)$ if
	$\alpha_1(f_1)= p\cdot q(x_1)$ for state $q$ of $A_1$ with $\iota_1(q)=f_1$.
\item	Assume that $\iota_1(q) = h^{(1)}$, and 
	$
	\trans{(h^{(1)},h^{(2)}),f, (h^{(1)}_1,h^{(2)}_1)\ldots (h^{(1)}_k,h^{(2)}_k)}
	$
	is a transition of $B$. Then $\trans{h^{(1)},f, h^{(1)}_1\ldots h^{(1)}_k}$ is a transition of $B_1$,
	and $A_1$ has a rule 
	\[
	q(f(x_1:h^{(1)}_1,\ldots,x_k:h^{(1)}_k)\to p\{x_j\mapsto q_j(x_j)\mid j\in J\}
	\]
	for some pattern $p$.
	For that, $A'_1$ receives the rule
	\begin{multline*}
	q(f(x_1:(h^{(1)}_1,h^{(2)}_1),\ldots,x_k:(h^{(1)}_k,h^{(2)}_k))\to \\
	p\{x_j\mapsto \angl{q_j,h^{(2)}_j}(x_j)\mid j\in J\}.
	\end{multline*}
\end{itemize}
The transducer $A'_2$ for $A_2$ is constructed analogously.
The transducers $A'_i$ can be constructed from $A_i$ and $B$ in polynomial time 
where $A_i$ and $A'_i$ are equivalent.
By Theorem \ref{t:canonical} we can compute for each of the $A'_i$ in polynomial time
the canonical earliest normal form as \DBTB\ (or \DLTB, respectively).
Since the advice automata of $A'_1$ and $A'_2$ coincide, $A'_1,A'_2$ (and thus also $A_1,A_2$)
are equivalent iff the normal forms of $A'_1$ and $A'_2$ are isomorphic. 
Since isomorphism of transducers is decidable in polynomial time, the corollary follows.
\end{proof}

\noindent
By Corollary \ref{c:poly-eq}, equivalence is polynomial-time for deterministic \emph{uc} transducers
with look-ahead, as well as for \emph{uc} transducers with inspection.

\section{How to Remove Look-ahead}\label{s:lookahead}

In the following, we assume that we are given a \emph{uc}- (or linear)
transducer $A$ with regular look-ahead, i.e., bottom-up deterministic advice automaton $B$. 
By Theorem~\ref{t:canonical}, we may assume that $A$ is canonical earliest. 
As usual, we assume that $A$ is given by $(Q,B,\iota,R,\alpha)$ where $B=(H,\Sigma,\delta,F)$.
Our goal is to decide whether the translation of $A$ can be realized by 
a \emph{uc}- (or linear) transducer \emph{without} look-ahead, but possibly with inspection.
A necessary condition for the latter is that the domain of the translation of $A$
can be accepted by a top-down deterministic automaton,
see Theorem~3.1 of~\cite{DBLP:journals/mst/Engelfriet77}.
By assumption, the domain of the translation of $A$ is given by the 
set $\L(B)$ of all trees accepted by $B$. If the translation can be realized by a 
\emph{uc}-transducer with inspection only, $\L(B) = \L(B')$ for some
top-down deterministic TA $B'$. One such $B'$ can be obtained by means
of the \emph{powerset} construction. The set $H'$ of states of $B'$ are 
non-empty subsets of states of $B$ where the set $F\subseteq H$ of accepting states of $B$ 
is the single accepting state of $B'$. 
Moreover, if $S\subseteq H$ is a state in $H'$, then for every input symbol $f\in\Sigma_k, k\geq 0,$ and
every $j\in\{1,\ldots,k\}$, 
\begin{multline*}
S_j =\{h_j\in H\mid\exists h\in S,\;h_1,\ldots,h_{j-1},h_{j+1},\ldots,h_k\in H\text{ such that }\\
	h = \delta(f,h_1\ldots h_k)\}\;\in\;H'
\end{multline*}
and $\angl{S,f,S_1\ldots S_k}$ is in the transition relation of $B'$.
For later use, we denote the top-down deterministic TA $B'$ constructed in this way from
$B$, by $B^\top$.
As $B$ is assumed to be trim, the automaton $B^\top$ is trim as well.
Checking whether or not $\L(B^\top) = \L(B)$ is decidable.
In fact, the two automata are equivalent iff for each 
transition $\angl{S,f,S_1\ldots S_k}$ constructed for $B^\top$, and every 
tuple of states $(h_1,\dots, h_k)\in S_1\times\ldots\times S_k$ there is some $h\in S$
such that $\trans{h,f,h_1\ldots h_k}$ is a transition of $B$.
This implies that it can be decided in PSPACE whether a 
bottom-up deterministic TA accepts a top-down deterministic language.
In fact, a \emph{polynomial} time algorithm for this task has recently been provided \cite{DBLP:journals/ipl/ManethS24}.

\subsection{The Construction}\label{ss:candidate}

Let us thus assume that $B$ and $B^\top$ are equivalent.
The following construction is given for uniform-copying transducers
(the construction for linear transducers is exactly the same where $\bigsqcup$ must be replaced with $\bigsqcup^{(1)}$, and
the notion of  ``common prefix'' in the construction below must be replaced with ``common 1-prefix'').
Let us assume that the transducer $A$ is equivalent to some \DBTI
transducer $\bar A$.
%
Then we construct a \DBTI transducer $A'$ as follows. 
Intuitively, the transducer $A'$ simulates the transducer $A$ for all possible look-ahead states and
only outputs the largest common prefix of all respective outputs.
In its state, it maintains for each look-ahead state $h$ the output that $A$ would have already produced 
but which cannot be produced by $A'$ yet.
The construction is essentially the same as the one given in \cite{DBLP:journals/tcs/EngelfrietMS16} for total transducers
(but without proof of termination).
Accordingly, each state of $A'$ is given by $\angl{\rho}$ for some mapping $\rho$ which assigns to the
states $h$ in some set $S\in H'$, trees $\rho(h)$ which are either in $\T_\Delta$
or of the form $\rho(h) = s\cdot q(x_1)$ where $s\in\T_\Delta(x_1)$ and
$q\in Q$ is a state of $A$ with $\iota(q) = h$.
For that state $\angl{\rho}$, we define $\iota'(\angl{\rho}) = S$.
Let 
\begin{multline*}
P_0 = \{\alpha(h)\in\T_\Delta\mid h\in F\}\cup\{p\in\T_\Delta(x_1)\mid\exists h\in F,q\in Q~\text{such that}\\
\alpha(h)= p\cdot q(x_1)\}.
\end{multline*}
The axiom $\alpha'$ of $A'$ is defined as follows. Let $s = \bigsqcup P_0$.
If $s\in\T_\Delta$, then $\alpha' = \{F\mapsto s\}$;
in this case, the set of states of $A'$ is empty.
Otherwise, $\alpha' = \{F\mapsto s\cdot\angl{\rho}(x_1)\}$, where
$\rho$ is given by $\rho(h) = u_h$ whenever $\alpha(h) = s\cdot u_h$.
In that case, we include $\angl{\rho}$ in the set $Q'$ of states of $A'$
with $\iota'(\angl{\rho}) = F$.

Assume now that $\angl{\rho}$ is a state of $A'$ with $\iota'(\angl{\rho}) = S$. 
Consider some input symbol $f\in\Sigma$ of rank $k\geq 0$ where
$\angl{S,f,S_1\ldots S_k}$ is a transition of $B^\top$.
In particular, $S_i\neq\emptyset$ for all $i$.
Let
\[
P=\{\rho(h)\in\T_\Delta\mid h\in S\}\cup
\{s_h\cdot p_h\mid \rho(h) = s_h\cdot q_h(x_1) \}
\]
where $A$ has a rule
\begin{equation}
q_h(f(x_1:h_1,\ldots,x_k:h_k))\to p_h\{x_j\mapsto q_{h,j}(x_j)\mid j\in J_h\}
\label{def:h-rule}
\end{equation}
for suitable $J_h\subseteq\{1,\ldots,k\}$, $p_h\in\T_\Delta(X_{J_h})$, and $q_{h,j}\in Q$.
The set $P$ consists of all patterns which $A'$ may attempt to output.
Let $p\in\T_\Delta(X_{J})$ denote the (uniquely defined) common prefix of all trees in $P$
for which there is a collection of trees $u_{j,h_j}, j\in J, h_j\in S_j$, of minimal sizes
with the following properties:
\begin{enumerate}
\item	For each $h\in S$ where $\rho(h)$ is ground,
\[
	\rho(h) = p\{x_j\mapsto u_{j,h_j}\mid j\in J\}
\]
	whenever $\trans{h,f,h_1\ldots h_k}$ is a transition of $B$;
\item	For each $h\in S$ where
	$\rho(h)$ is of the form $s_h\cdot q_h(x_1)$ and
	each rule of the form \eqref{def:h-rule},
\[
	s_h\cdot p_h\{x_j\mapsto q_{h,j}(x_j)\mid j\in J_h\} =
	p\{x_j\mapsto u_{j,h_j}\mid j\in J\}.
\]
%
\item 	For each $j\in J$,
	$u_{j,h_j}$ either is ground or contains the variable $x_j$.
\end{enumerate}
Assuming that $\bar A$ exists, we claim that the following hypothesis holds

\smallskip
\begin{description}
\item[\textbf{(H)}]
		The pattern $p$ and the collection of trees $u_{j,h_j},j\in J,h_j\in S_j$,
		exist and are uniquely determined.
\end{description}

\medskip

\noindent
Given that the hypothesis \textbf{(H)} holds,
we define for each $j\in J$, the mapping $\rho_j$ by
$\rho_j(h_j) = u_{j,h_j}$ ($h_j\in S_j$);
%
%
for $j\in J$, we add $\angl{\rho_j}$  to the set of states of $A'$ with $\iota'(\angl{\rho_j}) = S_j$, 
and introduce the rule
\begin{equation}
\angl{\rho}(f(x_1:S_1,\ldots,x_k:S_k))\to p\{x_j\mapsto\angl{\rho_j}(x_j)\mid j\in J\}.
\label{def:det_rule}
\end{equation}
This completes the construction.
We  illustrate the perhaps most complicated case by an example, namely, when 
all outputs stored in $\rho$ are ground. 

\begin{example}\label{e:pattern}
Assume that 
\[
\rho = \{h_1\mapsto f(a,g(c)), h_2\mapsto f(b,g(c)), 
	 h_3\mapsto f(a,b), h_4\mapsto f(b,b), h_5\mapsto c\}
\]
and for the binary input symbol $f$, $B$ has the transitions
\[
\angl{h_1,f,h_a h_c},
\langle h_2,f,
h_b h_c\rangle,
\angl{h_3,f,h_a h_b},
\angl{h_4,f,h_b h_b}
\]
while there is no transition for $f$ resulting in state $h_5$.
By comparing the outputs for $h_1$ and $h_2$, we identify
the subtrees $a$ and $b$ whose outputs cannot be decided depending on the input symbol
$f$ alone, but require information about the first child of $f$. 
Likewise, by comparing the outputs for $h_1$ and $h_3$, we identify
the corresponding subtrees $g(c)$ and $b$ whose outputs can be discriminated only 
depending on the second child of $f$ in the input.
Accordingly, the pattern $p$ is given by $p = f(x_1,x_2)$ where
$\rho_1=\{h_a\mapsto a,h_b\mapsto b\}$ and
$\rho_2=\{h_c\mapsto g(c),h_b\mapsto b\}$.
\qed
\end{example}

\noindent
%
%
If the construction succeeds, i.e., (1) never fails to construct for every attained state $\angl{\rho}$ and matching
transition of $B^\top$ the required rule, and (2) introduces only finitely many states $\angl{\rho'}$, then
the resulting transducer $A'$ is a \DBTI transducer.
Now assume that the construction has succeeded.
The following two lemmas summarize important properties of the resulting transducer $A'$.

\begin{lemma}\label{l:canon_ground}
Consider a state $\angl{\rho}$ of $A'$ for some mapping $\rho$ with domain $S$, $h\in S$ and $t\in\dom_B(h)$.
\begin{enumerate}
\item	If $\rho(h) = u$ is ground, then
	$\sem{t}_{A',\angl{\rho}}= u$;
\item	If $\rho(h) = u\cdot q(x_1)$ for some $u\in\T_\Delta(x_1)$, 
	then $\sem{t}_{A,q} = u\cdot\sem{t}_{A',\angl{\rho}}$.
	\ignore{
	and $q:t\to s$, then
	$\angl{\rho}:t\to u\cdot s$.
	}
	\qed
\end{enumerate}
\end{lemma}
\begin{lemma}\label{l:canon_context}
	Assume that $S\subseteq H$ is a state of $B^\top$, 
	and $t\in\C_\Sigma$ is a context with $(F,S):t$.
	Let $\sem{t}_A^{h} = s_{h}$ for $h\in S$.
	\begin{enumerate}
	\item	If $S =\{h\}$, i.e., just one look-ahead state is possible,
		then one of the following holds.
		\begin{itemize}
		\item If $s_h$ is ground, then $\sem{t}_{A'}^S = s_h$;
		\item If $s_h = u\cdot q(x_1)$, then
		$\sem{t}_{A'}^S = u\cdot\angl{\rho}(x_1)$ where
		$\rho = \{h\mapsto q(x_1)\}$.
		\end{itemize}
	\item	If $S$ contains more than one state of $S$, then one of the
		following holds.
	\begin{itemize}
	\item	All $s_h$ ($h\in S$) equal the same ground tree $u$.
		Then $\sem{t}_{A'}^{S} = u$;
	\item	Not all $s_h$ are equal.
		Let $u = \bigsqcup \{s_{h}\mid h\in S\}$ equal the maximal common prefix of 
		the $s_{h}$. Then 
		$\sem{t}_{A'}^{S} = u\cdot\angl{\rho}(x_1)$ where for all $h\in S$,
		$u\cdot\rho(h) = s_{h}$. \qed
	\end{itemize}
	\end{enumerate}
\end{lemma}

\noindent
The proofs of these two lemmas are by induction on the structure of $t$,
following the definition of~$A'$.

From Lemma \ref{l:canon_ground} we conclude that, upon successful termination, $A$ and $A'$ are equivalent.
Let $t\in\T_\Sigma$ with $h_0:t$ for some $h_0\in F$. 
%
%
First assume that $\alpha(h_0)$ is ground. Then the output produced by $A$ for $t$ is $\alpha(h_0)$.
If $\alpha'(F)$ is ground, then $\alpha'(F)=\alpha(h_0)$ by Lemma \ref{l:canon_ground} 
and therefore,
\[
\sem{t}_{A'} = \alpha'(F) = \alpha(h_0) = \sem{t}_A.
\]
Otherwise by construction of the axiom of $A'$, 
$\alpha'(F) = u\cdot\angl{\rho}(x_1)$ holds for some $\rho$ where $\alpha(h_0) = u\cdot\rho(h_0)$ and $\alpha(h_0) = u\cdot\rho(h_0)$.
Altogether therefore,
\[
\sem{t}_{A'} = u\cdot\rho(h_0) = \alpha(h_0) = \sem{t}_A.
\]
Now assume that $\alpha(h_0) = u\cdot q_0(x_1)$ for some $q_0\in Q$ with $\iota(q_0) = h_0$.
%
Again by construction of the axiom of $A'$, $\alpha'(F) = u_1\cdot\angl{\rho}(x_1)$ for some prefix $u_1$ of $u$ and state $\rho$
such that $\rho(h_0) = u_2\cdot q_0(x_1)$ with $u = u_1\cdot u_2$.
Then by Lemma \ref{l:canon_ground}, $\sem{t}_{A',\angl{\rho}} = u_2\cdot\sem{t}_{A,q_0}$. 
Therefore,
\[
\sem{t}_{A'} = u_1\cdot\sem{t}_{A',\angl{\rho}} = u_1\cdot u_2\cdot\sem{t}_{A,q_0} = \sem{t}_A.
\]
\ignore{
%
%
For the reverse direction, consider the axiom $\alpha'(F)$ of $A'$.
If it is ground then, as argued above, all axioms of $A$ are ground as well and equal $\alpha'(F)$ -- implying that
the transducers are equivalent.
Otherwise, $\alpha'(F) = u_1\cdot\angl{\rho}(x_1)$ for some $\rho$
where $\alpha(h_0) = u_1\cdot\rho(h_0)$.
Let $\angl{\rho}:t\to s$ denote the computation of $A'$ for $t$ and $s$ the corresponding output tree.
If $\alpha(h_0) = u_1\cdot u_2$ is ground, then $\rho(h_0) = u_2$ is ground and therefore by Lemma \ref{l:canon_ground}.1, 
$\angl{\rho}:t\to u_2$ holds.
If on the other hand, $\alpha(h_0) = u_1\cdot u'_2\cdot q_0(x_1)$ for some state $q_0$ with
$\iota(q_0) = h_0$, then by Lemma \ref{l:canon_ground}.2, $q_0: t\to s'$ where $s = u'_2\cdot s'$ holds.
}
Altogether, the outputs of $A$ and $A'$ for $t$ coincide.
\qed
\subsection{Success}\label{ss:success}
Assuming that the \DBTR $A$ is equivalent to the \DBTI $\bar A$, we now show that 
our hypothesis \textbf{(H)} from the previous subsection holds whenever a new rule is constructed.
For that, we introduce the following auxiliary lemma which inductively relates the states $\bar q$ of $\bar A$
with corresponding state of $A'$.

\begin{lemma}\label{l:aux}
%
%
Assume that $\L(B)$ is top-down deterministic, and let $B^\top$ be the corresponding deterministic top-down
tree automaton obtained from $B$ by the powerset construction.
Let $\bar A=(B^\top,\bar Q,\bar \iota,\bar R,\bar \alpha)$ be an earliest \emph{uc} transducer 
which is equivalent to $A$, and uses $B^\top$ for inspection.
Then there is a mapping $\pi$ which maps each state $\bar q\in\bar Q$ of $\bar A$ 
with $\bar\iota(\bar q) = S$,
to some $\pi(\bar q)$ which maps each $h\in S$, either to a ground tree in $\T_\Delta$ or to some
tree of the form $u\cdot q(x_1)$ where $u\in\T_\Delta(x_1)$ and $q\in Q$ such that the following holds:
\begin{enumerate}
\item	If $\bar\alpha(F)\in\T_\Delta$ then $\bar\alpha(F) = \alpha'(F)$;
	if $\bar\alpha(F) = u\cdot \bar q(x_1)$, then $\alpha'(F) = u\cdot\angl{\pi(\bar q)}(x_1)$.
\item	Assume that $\bar q\in\bar Q$, and $\angl{\pi(\bar q)}$ has been constructed as a state of $A'$, and
	\[
	\bar q(f(x_1:S_1,\ldots,x_k:S_k))\to p\{x_j\mapsto \bar q_j(x_j)\mid j\in J\}
	\]
	is a transition of $\bar A$ for $f\in\Sigma^{(k)}$ and
	for some $p\in \T_\Delta(X_{J})$ and $\bar q_j\in\bar Q$.
	Then the pattern $p$ together with the family of trees
	$u_{j,h_j} = \pi(\bar q_j)(h_j), j\in J, h_j\in S_j$, satisfy the hypothesis~\textbf{(H)}.
	Accordingly, $A'$ receives the states $\angl{\pi(\bar q_j)}, j\in J$, together with the transition 
	\[
	\angl{\pi(\bar q)}(f(x_1:S_1,\ldots,x_k:S_k))\to p\{x_j\mapsto\angl{\pi(\bar q_j)}(x_j)\mid j\in J\}.
	\]
\end{enumerate}
\end{lemma}

\noindent
By Item 1, the axioms of $\bar A$ and $A'$ are related by the mapping $\pi$, whereas Item 2 corresponds to the inductive step:
if the left-hand side of a rule of $\bar A$ is mapped to some state of $A'$ via $\pi$ then there is a matching rule (with identical
pattern $p$) also provided by $A'$ so that the states $\bar q_j$ of $\bar A$ are mapped to corresponding states of $A'$ via $\pi$.
In this way, the transducer $\bar A$ is homomorphically mapped onto the transducer $A'$.

\ignore{
such that for every context $c\in\C_\Sigma$ with $(F,S):c$ in $B^\top$,
the following holds:

	If $\alpha'(F) = s'_0\cdot q'_0$ and $q'_0,c \to s'_1\cdot q'(x_1)$, 
	then for every $h\in S$ with $(h_0,h):c$ for some $h_0\in F$, it holds that
	$\alpha(h_0)$ is of the form $s_0\cdot q_0(x_1)$ so that
	$q_0,c\to s_1\cdot q(x_1)$ for some state $q$ of $A$ with $\iota(q) = h$ such that
	$s_0\cdot s_1 = s'_0\cdot s'_1\cdot \phi(h)$.
}

\begin{proof}
The \emph{uc}-transducer $\bar A$ with inspection $B^\top$ can be considered as a 
\emph{uc}-transducer with look-ahead $B$ which simply ignores the look-ahead.
Technically, we construct the \emph{uc}-transducer
$\bar A\otimes B$ with look-ahead $B$ as the tuple $(B,\bar Q_B,\bar\iota_B,\bar R_B, \bar\alpha_B)$ where
\begin{itemize}
\item	$\bar Q_B = \{\angl{\bar q,h}\mid \bar q\in \bar Q, h\in\bar \iota(\bar q)\}$ with
	$\bar\iota_B:\bar Q_B\to H$ given by $\bar\iota_B(\angl{\bar q,h}) = h$;
\item	$\bar R_B$ consists of all rules 
\[
\angl{\bar q,h}(f(x_1:h_1,\ldots,x_k:h_k))\to p\{x_j\mapsto\angl{\bar q_j,h_j}(x_j)\mid j\in J\}
\]
where there is a rule $\bar q(f(x_1:S_1,\ldots,x_k:S_k))\to p\{x_j\mapsto\bar q_j(x_j)\mid j\in J\}$ in $\bar R$ 
with $h\in\iota(q), h_1\in S_1,\ldots,h_k\in S_k$, $\trans{h,f,h_1\ldots h_k} \in\delta$, and some pattern $p\in\T_\Delta(X_J)$
for some $J\subseteq\{1,\ldots,k\}$;
and finally,
\item	The mapping $\bar\alpha_B$ is obtained from $\alpha$ by defining for $h\in F$,
	$\bar\alpha_B(h) = \bar\alpha(F)$ if $\bar\alpha(F)\in\T_\Delta$, and
	$\bar\alpha_B(h) = u\cdot\angl{\bar q,h}(x_1)$ if $\bar\alpha(F) = u\cdot\bar q(x_1)$.
\end{itemize}
By construction, $\bar A\otimes B$ and $\bar A$ are equivalent. 
%
%
We construct the mapping $\pi:\bar Q\to Q'$ by means of the aheadness map of $\bar A\otimes B$ as
\[
\pi(\bar q) = \{ h\mapsto \textsf{ah}(\bar q,h)\mid h\in \iota(\bar q)\}
\]
\ignore{
Let $\bar q\in\bar Q$ and $h\in\iota(\bar q)$.
Then $(\bar q,h)$ is a state of $\bar A\otimes B$.
First assume that $(\bar q,h)$ is a constant state, i.e., 
there is a ground tree $s$ such that $\sem{t}_{\bar A,\bar q} = s$ for all $t\in\dom_B(h)$. 
In this case, we set $\pi(\bar q)(h) = s$.
If $(\bar q,h)$ is a non-constant state of $\bar A\otimes B$, we can apply the 
aheadness mapping \textsf{ah} to obtain how much $A$ is ahead of $\bar A\otimes B$.
In this case, we therefore define
$\pi(\bar q)(h) =  \textsf{ah}(\bar q,h)$.
}
For every context $t\in\C_\Sigma$ reaching the state $\bar q$ with $\iota(\bar q) = S$, the mapping $\pi$ thus provides for 
every $h\in\iota(\bar q)$,
the output \emph{retained} for look-ahead $h$, i.e., 
if $\sem{t}_{\bar A}^S = u\cdot\bar q(x_1)$ and $\pi(\bar q)(h) = v$, then 
%
%
according to Lemma \ref{l:canon_context}, the tree $v$ equals $\rho(h)$ where $\sem{t}_{A'}^S = u\cdot\angl{\rho}(x_1)$ holds.
Therefore, the mapping $\pi$ satisfies the properties of the lemma.
\end{proof}

\noindent
Lemma \ref{l:aux} not only provides us with the insight that the construction of $A'$ succeeds whenever there
is a \DBTI~transducer $\bar A$ (with inspection automaton $B^\top$) equivalent to $A$, but also proves that $A'$ in fact represents 
the \emph{canonical} earliest \DBTI~ transducer with this property.

\begin{corollary}\label{c:barprime}
Assume that $\bar A$ is a \emph{uc}-transducer with inspection $B^\top$,
which is equivalent to $A$. Then the transducer $A'$ construction in Section \ref{ss:candidate}, is the canonical earliest 
\emph{uc}-transducer with inspection $B^\top$ which is equivalent to $A$.
\qed
\end{corollary}

\subsection{Termination}\label{ss:term}
It remains to prove that the addition of further states $\angl{\rho}$ and rules to $A'$ eventually terminates.
%
We prove that the depths of trees recorded in the mappings $\rho$ remains bounded -- whenever 
there is a \DBTI~transducer $\bar A$ equivalent to $A$.
Our proof is based on the following three properties. 

\begin{description}
\item[\textsf{Synchronizing.}]
The \DBTR transducer $A$ is \emph{synchronizing}, if 
for every context $t\in\C_\Sigma$, input symbol $f\in\Sigma$ of rank $k\geq 2$, 
and states $q_1,q_2$ with $\iota(q_i) = h_i$
where $\sem{t}_{A}^{h_i} = s_i\cdot q_i(x_1)$ for $i=1,2$,
the following holds.
Let $q_i(f(x_1:h_{i,1},\ldots,x_k:h_{i,k}))\to T_i$ be rules of $A$ according to 
\eqref{eq:rule}
%
and assume that there are indices $j_1\neq j_2$ so that both $x_{j_1}$ and $x_{j_2}$ occur in $T_1$ or in $T_2$.
\ignore{
$h_{1,j_i}\neq h_{2,j_i}$ for $i=1,2$.
}
Then there are 
factorizations $T_i = v_i\cdot p\{x_j\mapsto u_{i,j}\mid j\in J\}$ , $i=1,2$, for 
$v_i\in\T_\Delta(x_1)$, the same pattern $p\in\T_\Delta(X_J)$ for
some set $X_J$ of variables where 
each $u_{i,j}$ is either ground, or is of the form $u'\cdot q'(x_j)$ for suitable $u'$, $q'$,
such that 
\[
	s_1\cdot v_1 = s_2\cdot v_2.
\]
\ignore{
\begin{itemize}
\item	$s_1\cdot p_1 = s_2\cdot p_2$; or
\item	$J=\{j\}$ for some index $j$.
\end{itemize}
}
\item[\textsf{Twinning}.]\quad
The \DBTR transducer $A$ is \ignore{(monadicly)} \emph{twinning}, 
if the following holds for 
states $q_1,q_2$ with $\iota(q_i) = h_i$, and contexts $t,t'\in\C_\Sigma$ such that
$\sem{t}_{A}^{h_i} = s_i\cdot q_i(x_1)$ and
$\sem{t'}_{A,q_i}^{h_i} = u_i\cdot q_i(x_1)$.
\begin{itemize}
\item	Either $u_1 = u_2 = x_1$,
\item	or there are trees $v,w\in\T_\Delta(x_1)$ such that
	$s_1 = s_2\cdot w$, $u_1 = v\cdot w$ and $u_2 = w\cdot v$ 
 	or vice versa,
	$s_2 = s_1\cdot w$, $u_2 = v\cdot w$ and $u_1 = w\cdot v$.
\end{itemize}
\item[\textsf{Erasing}.]\quad
The \DBTR transducer $A$ is \emph{erasing}, 
if the following holds for all contexts $t,t'\in\C_\Sigma$, 
and states $q_1,q_2$ with $\iota(q_i) = h_i$. 
Assume that $\sem{t}_{A}^{h_i} = s_i$ for $i=1,2$ where $s_1$ is of the form $u_1\cdot q(x_1)$ 
and $s_2$ is ground. 
Then $\sem{t'}_{A,q}^{h_1}= u\cdot q(x_1)$ for some $u\in\T_\Delta(x_1)$ and $(h_2,h_2):t'$,
implies that $u=x_1$.
\end{description}
Intuitively, \emph{synchronization} means that at input symbols $f$ where the output may depend 
on the look-ahead for more than one child, the differences in the outputs so far must be leveled out.
This means that \emph{aheadness} in the output can only be accumulated when differences can be attributed to a single child only.
%
%
The second property \emph{twinning} on the other hand, refers exactly to the latter situation. 
It is the straightforward generalization
of the corresponding property for string transducers as considered in \cite{DBLP:journals/tcs/Choffrut77,DBLP:conf/icalp/ReynierV19}.
It indicates that the outputs during a joint repetition between a pair of states does not affect the \emph{aheadness}.
%
%
The third property \emph{erasing} means that if for a pair of jointly possible look-ahead states, the output for one is a ground tree,
then a subsequent iteration of the transducer with the other state cannot produce 
arbitrarily large outputs.
We remark that the latter property as well as \emph{synchronization} have no 
analogues in the string case.
%

\newcommand{\midd}{\mid\!\mid}
\renewcommand{\middle}[1]{\midd\!\! #1 \!\!\midd}
The \emph{variation} $\middle{s_1,s_2}$ of trees $s_i$ where for each $i$, $s_i\in\T_\Delta$ or of the form $s_i = s'\cdot q'(x_1)$
with $s'\in\T_\Delta(x_1)$, and not for both $i=1,2$, $s_i\in\T_\Delta$ is defined as follows.
If $s_1 = s_2$ then $\middle{s_1,s_2} = 0$. Otherwise,
let $s_0\in\T_\Delta(x_1)$ be maximal so that $s_i = s_0\cdot u_i, i=1,2$ for some trees $u_i,i=1,2$.
Then $\middle{s_1,s_2}$ equals the maximal depth of trees $u_1,u_2$.
%
%
Thus, e.g., 
$\middle{g(a, q(x_1)), g(a,a)} = 1$, and
$\middle{g(a, q(x_1)), g(b,q'(x_1))} = 2$ for some other state $q'$. 
%
%
Here, both leaves $a\in\Delta$ as well as expressions $q(x_i)$ for states $q$ and variables $x_i$
are assumed to have depth 1.

The \DBTR transducer $A$ has \emph{bounded variation} if there is some $K\geq 0$ such that for every 
$t\in\C_\Sigma$,  
and states $q_i$, with $\iota(q_i) = h_i$ and $\sem{t}_{A}^{h_i} = s_i$ for $i=1,2$, where at least one $s_i$ is non-ground, 
$\middle{s_1,s_2}\leq K$ holds.

\medskip
Now assume that $A$ is synchronizing, erasing and twinning. 
Then the outputs of
any two computations for the same input tree, cannot differ much. 
Intuitively, the variation is synchronized at pairs of rules mentioning more than one $x_j$, 
does not increase in joint pairs of ``monadic loops''s and 
may increase only marginally once one of the outputs is ground.
%
%
%
Altogether, we prove:

\begin{lemma}\label{l:variation}
Assume that 
the bottom-up deterministic look-ahead automaton $B$ of the \emph{uc}-transducer $A$
has $m\geq 1$ states where $\L(B)$ is a top-down deterministic language where
$B^\top$ is the top-down deterministic FA corresponding to $B$.
\begin{enumerate}
\item If $A$ is equivalent to some \DBTI $\bar A$ with inspection $B^\top$, then
	$A$ is synchronizing, erasing, and twinning.
\item If the \DBTR\ $A$ is synchronizing, erasing, and twinning, 
then $A$ has bounded variation where the bound is given by $|A|\cdot(|A|+m)$. 
\end{enumerate}
\end{lemma}

\begin{proof}
The proof that $A$ then must be twinning follows along the same lines as for string transducers.
Here, we only consider synchronization and erasure. 
%
%
Assume that $A$ is equivalent to some \DBTI\ $\bar A$
with inspection automaton $B^\top$. 
By Corollary \ref{c:barprime}, we may assume w.l.o.g.\ that $\bar A$ equals $A'$ as constructed in Subsection \ref{ss:candidate}.
Let $t\in\C_\Sigma$ denote a context, 
and assume 
that $q_1,q_2$ are states with $\iota(q_i) = h_i$,
so that $\sem{t}_{A}^{h_i} = s_i\cdot q_i(x_1)$ for $i=1,2$.
Then $\sem{t}_{A'}^S = u\cdot\angl{\rho}(x_1)$ holds for some set $S$ where $h_1,h_2\in S$
with
$u\cdot\rho(h'_i) = s_i$ for $i=1,2$.
Now consider some input symbol $f\in\Sigma$ of rank $k\geq 2$, and rules
$q_i(f(x_1:h_{i,1},\ldots,x_k:h_{i,k}))\to T_i$ 
of $A$ according to
\eqref{eq:rule}
%
where there are indices $j_1\neq j_2$ so that both $x_{j_1}$ and $x_{j_2}$ occur in $p_1$ or $p_2$.
By construction, $A'$ must have a rule
\[
\angl{\rho}(f(x_1:S_1,\ldots,x_k:S_k))\to p\{x_j\mapsto\angl{\rho_j}(x_j)\mid j\in J\}.
\]
so that 
\[
\rho(h'_i)\cdot T_i = p\{x_j\mapsto \rho_j(h_{i,j})\{x_1\mapsto x_j\}\mid j\in J\}
\]
holds for $i=1,2$. Since $x_{j_i}$ occurs either in $p_1$ or in $p_2$, $x_{j_i}$ must also occur in $p$.
Let $p'$ be the minimal subpattern of $p$ which contains all occurrences of variables so that $p = v\cdot p'$.
Since $\rho(h_i)$ is also a prefix of $p$, we have $v = \rho(h_i) \cdot u_i$.
Since 
\[
\begin{array}{lll}
&&p\{x_j\mapsto \rho_j(h_{i,j})\{x_1\mapsto x_j\}\mid j\in J\}\\
&=&
v\cdot p'\{x_j\mapsto \rho_j(h_{i,j})\{x_1\mapsto x_j\}\mid j\in J\} 	\\
&=&
\rho_i(h_i)\cdot u_i\cdot p'\{x_j\mapsto \rho_j(h_{i,j})\{x_1\mapsto x_j\}\mid j\in J\} 	\\
&=&
\rho_i(h_i)\cdot T_i
\end{array}
\]
we conclude by top cancellation, that
$u_i$ is a prefix of $T_i$, i.e., $T_i = u_i\cdot T'_i$ for suitable $T'_i$
where
$\rho(h'_1)\cdot u_1 = \rho(h'_2)\cdot u_2$ holds. Accordingly, we have that
\[
s_1\cdot u_1 = u\cdot\rho(h'_1)\cdot u_1 = u\cdot\rho(h'_2)\cdot u_2 = s_2\cdot u_2
\]
holds, and the claim follows.

Now consider the property of being erasing.
Assume that $\sem{t}_{A}^{h_i} = s_i$ for $i=1,2$ where $s_2$ is ground, and $s_1 = u_1\cdot q(x_1)$ for some 
pattern $u_1\in\T_\Delta(x_1)$ where $\iota(q) = h_1$.
Assume further that $\sem{t'}_{A,q}^{h_1} = v'\cdot q(x_1)$ where $(h_2,h_2):t'$ holds.
Let $t_i\in\T_\Sigma$ denote some input tree in $\dom_B(h_i)$ and for $k\geq 0$, define 
$t_k^{(i)}$ as the tree $t\cdot (t')^k\cdot t_i$. Let $\sem{t_1}_{A,q_1} = v\in\T_\Delta$.
Then for all $k\geq 0$, $\sem{t_k^{(1)}}_{A} = u_1\cdot (v')^k\cdot v$ 
while $\sem{t_k^{(2)}}_{A} = s_2$. 

We claim that then $v' = x_1$. For a proof, consider the set $S$ of states so that
$(F,S):t$ holds for $B^\top$.
Then $\sem{t}_{A'}^S$ equals some output tree $s$ of the form $u\cdot\angl{rho}(x_1)$ for some
mapping $\rho$ with domain $S$ where $u_1 = u\cdot u'_1$ and $s_2 = u\cdot s'_2$ with $\rho(h_1) = u'_1$
and $\rho(h_2) = s'_2$.
And we can define sequence $S_k, k\geq 0,$ as follows:
$S_0 = \iota'\angl{\rho}$ and for $k>0$, $S_k$ as the state of $B^\top$ with $(S_{k-1},S_k):t'$ in $B^\top$.
Accordingly, we also define the sequence $\rho_k,k\geq 0$ with $\iota'\angl{rho_k} = S_k$ by
$\rho_0 = \rho$, and for $k>0$, $\rho_k$ as the state of $A'$ with
$\sem{t'}_{A',\angl{\rho_{k-1}}}^{S_k} = v_k\cdot\angl{\rho_k}(x_1)$.

Assume for a contradiction that $v'$ were different from $x_1$, and
consider $\sem{t\cdot(t')^k}_{A'}^{(S_k)} = u\cdot v_1\cdot\ldots\cdot v_k\cdot\angl{\rho_k}(x_1)$.
All these outputs must also be prefixes of of $s_2 = u\cdot s'_2$ -- implying that there is some $k_0\geq 0$ 
such that $v_k = x_1$ for all $k\geq k_0$.
Since for all $k\geq 0$,
\[
\begin{array}{lll}
u_1 \cdot(v')^k	&=& 	\sem{t\cdot (t')^k}_{A}^{h_1} \\
		&=& 	u\cdot v_1\cdot\ldots\cdot v_k\cdot(\rho_k(h_1)) 
\end{array}
\]
this means that the sizes of trees retained in $\rho_k, k\geq k_0,$
for $h_1$ strictly increase, implying that the $\rho_k, k\geq k_0,$ all are distinct -- 
in contradiction to the finiteness of the number of states of $A'$.

It remains to prove the second assertion of Lemma \ref{l:variation},
namely, that every \DBTR\ $A$ which is synchronizing, erasing and twinning, 
has bounded variation where the bound is given by
$|A|\cdot(|A|+m)$.

Let $B^\top$ denote the top-down deterministic automaton corresponding to the
look-ahead automaton $B$ of $A$,
and assume for a contradiction that $t\in\C_\Sigma$ is a context with
a minimal number of nodes violating the claim of the lemma.
Let $S\subseteq H$ denote the state of $B^\top$ such that $(F,S):t$ holds. 
Assume that 
$t = t_1\cdot\ldots\cdot t_m$ where 
$t_i= f_i(u_{i,1},\ldots,u_{i,j_i-1},x_1,u_{i,j_i+1},\ldots, u_{i,k_i})$
for some some $1\leq j_i\leq k_i$, some $f_i\in\Sigma^{(k_i)}$ and ground trees $u_{i,j'},j'\neq j_i$.
Let $h_{i,0},\ldots,h_{i,m}$ denote states of $B$ so that
$(h_{i-1},h_i):t_i$ holds for $i=1,\ldots,m$
where $h_{i,0}\in F$, and $h_{i,m}\in S$.
Assume that $\sem{t}_{A}^{h_{i,m}} = s_i$,
$i=1,2$, holds.
If $m=0$, then $s_i = \alpha(h_{i,0})$ with a variation bounded by $|A|$ -- in contradiction to the assumption.
%
%
Otherwise, consider the maximal $m'\leq m$ where
$\sem{t_1\ldots t_{m'-1}}_{A}^{h_{i,m'}} = u_i\cdot q_i(x_1)$ holds for some patterns $u_i\in\T_\Delta(x_1)$ 
and states $q_i$ such that $\sem{t_{m'}}_{A,q_i}^{h_{i,m'}} = v_i\cdot p\sigma_i$ for some $p\in\T_\Delta(X_J)$ and
substitutions $\sigma_i$ such that the following holds:
\begin{itemize}
\item	$u_1\cdot v_1 = u_2\cdot v_2$; 
\item	$\sigma_1(x_{j}) = \sigma_2(x_j)\in\T_\Delta$ for $j\neq j_{m'}$;
\item	$\sigma_i(x_{j_{m'}})$ is a subtree of some right-hand side of $A$.
\end{itemize}
If $m'=m$, then the claim of the lemma holds for $t$ -- in contradiction to our assumption. 
Therefore, assume that 
$m'<m$. 
%
%
Assume that $v$ is obtained from $p$ 
by substituting $x_j$ with $\sigma_1(x_j)$ for all $j\neq j_{m'}$.
Thus, $s' = u_1\cdot v_1\cdot v$ is a common prefix of $s_1$ and $s_2$, while 
a difference occurs only at $w_i = \sigma_i\,x_{j_{m'}}$.
In case that the properties above are never jointly satisfied, 
we set $m'=0$, and $s'$ as the maximal common prefix of the axioms $\alpha(h_1), \alpha(h_2)$ in $\T_\Delta(x_1)$ with
$s' \cdot w_i = \alpha(h_i)$ for $i=1,2$.

Assume that $m''\geq m'$ is chosen maximal so that 
$\sem{t_{m'+1}\ldots t_{m''}}_{A,q_i}^{h_{i,m''}} = v'_i\cdot q'_i(x_i)$, i.e., both
outputs are non-ground. In particular, due to the synchronization property, each of the right-hand sides
contributing to the output $v'_i$ contains exactly one variable. 
The variation for the context $t_1\cdot\ldots\cdot t_{m''}$ then is given by
$\middle{w_1\cdot v'_1\cdot q'_1(x_1),w_2\cdot v'_2\cdot q'_2(x_1)}$.
%
%
Assume for a contradiction that this variation is not bounded by 
$|A|^2$.
Then due to the synchronization property and the choice of $m'$, this implies that 
$m''-m'> n^2$ where $n$ is the number of states of $A$. 
%
%
Therefore, at least one pair of states
occurs at least twice. But then, due to the twinning property, the same variation
is attained with a smaller context --- contradicting the minimality of $t$.

To continue with our argument, we conclude that $m''$ must be less than $m$.
By the definition of $m''$ this means that one of the right-hand sides
chosen for $q'_i$ and $f_{m''+1}$ must be ground.
W.l.o.g., assume that this is the right-hand side $T_1$ for $q'_1$. 
But then due to the erasing property, the depth of the output for 
$t_{m''+1}\cdot\ldots\cdot t_m$ is bounded by $|A|\cdot m$.
Altogether therefore, the variation is bounded by $|A|^2+ |A|\cdot m= |A|\cdot(|A|+m)$ ---
in contradiction to our assumption.
this concludes the proof of Lemma \ref{l:variation}.
\end{proof}

\noindent
In summary, we obtain the main theorem of this section:

\begin{theorem}\label{t:removal}
Let $A$ be a \DBTR (\DLTR). It is decidable whether or not the translation of $A$
can be realized by a \DBTI (\DLTI).
\end{theorem}


\begin{proof}[Proof of Theorem \ref{t:removal}]
By Theorem~\ref{t:canonical} we may assume that $A$ is earliest and canonical.
The procedure then proceeds as follows. 
First, the top-down deterministic automaton $B^\top$ is constructed corresponding to the look-ahead automaton $B$ of $A$.
If $B^\top$ is not equivalent to $B$, we know that the domain of the transducer is not top-down deterministic.
Therefore, we report \textsf{failure} and quit.

Next, we proceed with the construction of a transducer $A'$.
Assume that a state $\angl{\rho}$ is encountered where $\middle{\rho(h_1),\rho(h_2)}$ exceeds
$b=|A|\cdot(|A|+m)$ ($m$ the number of states of $B$). Then by Lemma \ref{l:canon_context}, 
$A$ cannot have variation $b$. Therefore by Lemma~\ref{l:variation}, $A$ cannot be equivalent to
a \DBTI. Accordingly, we report \textsf{failure} and quit.

Otherwise, only a bounded number of states $\angl{\rho}$ are encountered.
If the construction fails to construct one of the rules required for $A'$
due to violation of the hypothesis \textbf{(H)},
we conclude from Lemma \ref{l:aux} that no equivalent \DBTI\ exists; 
we therefore report \textsf{failure} and quit.

If on the other hand, the construction of all required rules of $A'$ succeeds, 
we report \textsf{success} and return the constructed transducer $A'$, which
by Lemma~\ref{l:canon_ground} 
and Corollary \ref{c:barprime} is the canonical earliest \DBTI~equivalent to $A$.
\end{proof}

\section{How to Inspect Top-Down Deterministic Languages}\label{s:check}\label{s:inspect}

Now consider a \DBTI $A$ with underlying top-down deterministic automaton $B$ which
is assumed to be canonical earliest.
%
%
For the following, we denote the unique state $h$ of $B$ with $\dom_B(h) = \T_\Sigma$
(given that there is such a state), by $\top$. 
The \DBTI (\DLTI) $A$ is \emph{without inspection} (denoted by \DBT and \DLT, respectively)
if for every rule $q(f(x_1:h_1,\ldots,x_k:h_k)\to T$ of $A$, $h_j=\top$ whenever 
$x_j$ does not occur in $T$.
In particular, this is the case when $A$ is \emph{non-deleting}, i.e., 
the right-hand side of every rule of $A$ contains all variables $x_j$ occurring in its left-hand side.

Any \DBTI (\DLTI) that is without inspection can be changed in such a way that no advice automaton
is present at all (see the discussion of various transducer models in Section \ref{s:basics}).
In order to understand the idea of our approach to inspection removal, 
consider the rule 
\[
q(f(x_1 : h_1 , x_2 : h_2 )) \to g(q_1(x_1) , r(b))
\]
of a \emph{uc}-transducer where $\dom_B (h_2) = \{g(a,t) \mid t \in \T_\Sigma\}$. 
Since $x_2$ does not occur in the right-hand side of the rule, the transducer will not visit the corresponding 
subtrees of the input. It therefore seems impossible for a transducer without inspection 
to verify that the corresponding subtrees are in $\dom_B(h_2)$.
However, due to the simple structure of the input trees in that language,
an artificial call could be introduced which successively checks the input while producing some constant 
output subtree of the given rule.
In our particular case, the given rule can be simulated by
\[
\begin{array}{lll}
q(f(x_1 , x_2 )) &\to&  g(q_1(x_1) , r(q_2(x_2)))\\
q_2(g(x_1,x_2))	 &\to&	q'_2(x_1)	\\
q'_2(a)		 &\to&	b
\end{array}
\]
For convenience, we have dropped the annotations of the $x_i$ with advice states.

To generalize this idea, we require an effective method to decide for a given top-down deterministic tree language $L$ and
a ground output tree $s\in\T_\Delta$ whether or not $L$ is \emph{output-recognizable} by $s$, i.e.,
whether a \emph{uc} (or, better, linear) transducer $A$ without inspection with domain $L$ can be constructed such that
$\sem{t}_A = s$ for all $t\in L$.
It turns out that a language is output-recognizable via some \DBT\ iff
it is output-recognizable via some \DLT. Hence, we drop the qualification.
%

Assume that the language $L$ is accepted by a trim
top-down deterministic TA.
We show that for a fixed input alphabet, it can be decided in \emph{polynomial time} whether or not $L$
is output-recognizable 
by a particular output tree $s$.
It turns out that output symbols in the output tree $s$ with arities exceeding one, vastly extend the potential for checking 
top-down deterministic languages $L$ via transducers without inspection.

\begin{example}\label{e:binary}
Consider the rule
\[
q(f(x_1:h_1,x_2:h_2))\to g(g(c,c), q(x_2))
\]
where $L= \dom_B(h_1)$ equals the set of trees $t$ over $\Sigma=\{a,b,f\}$ for nullary symbols $a,b$ and binary symbol $f$
so that $t$ has at least two leaves where the leftmost leaf of $t$ equals $a$ and the rightmost leaf equals $b$.
The language $L$ is \emph{not} output-recognizable by
a single leaf $c$ alone, because both the leftmost and the rightmost leaf of the input must be visited by the transducer,
and a single output leaf only allows visiting a single path in the input tree.
On the contrary, the output tree $g(c,c)$ offers the possibility to visit two paths in the input thereby allowing to recognize $L$.
A corresponding transducer is given by the following rules:
\[
\begin{array}{lll}
q(f(x_1,x_2))	&\to&	 g(q_0(x_1), q(x_2))	\\
q_0(f(x_1,x_2))	&\to&	 g(q_a(x_1), q_b(x_2))	\\
q_a(f(x_1,x_2))	&\to&	 q_a(x_1)		\\
q_a(a)	&\to&	 c	\\
q_b(f(x_1,x_2))	&\to&	 q_b(x_2)		\\
q_b(b)	&\to&	 c\,.	\\
\end{array}
\]
\qed
\end{example}
%
\ignore{
We recall the following definition.
A directed graph $G=(V, E)$ is \emph{strongly connected} if for every $v_1, v_2\in V$ there is a path
from $v_1$ to $v_2$. 
A subgraph $G'$ of $G$, i.e., $G'=(V',E')$ such that $V'\subseteq V$ and $E'\subseteq E$, is a \emph{strongly connected component}
if it is strongly connected.
}
Consider a top-down deterministic TA $B=(H,\Sigma,\delta,\{h_0\})$. 
\ignore{
The set $H$ of states and the transition relation $\delta$ induce a directed graph $G_B=(V_B,E_B)$ where $V_B=H$ and
$E_B$ is induced by $\delta$ as follows:
If $\trans{h,f,h_1\ldots h_k}\in \delta$ then $(h,h_j)\in E$ for $j=1,\ldots,k$.
Then the following holds.
\begin{lemma}\label{l:output-recognizable}
	Let the language $L$ be accepted by the trim top-down deterministic TA $B$ and $\Delta$ an output alphabet 
	with at least one element of rank 0. 
\begin{itemize}
\item 	If the maximal rank of output symbols in $\Delta$ is bounded by 1,
	$L$ is output recognizable w.r.t.\ $\Delta$ iff for every transition $\trans{h,f,h_1\ldots h_k}$ of $B$,
	$h_j\neq\top$ for at most one index $j\in\{1,\ldots,k\}$;
\item	
	Let $G_B$ denote the directed graph induced by $B$.
	If $\Delta$ contains at least on symbol of rank exceeding 1,
	$L$ is output-recognizable w.r.t.\ $\Delta$ iff for every transition $\trans{h,f,h_1\ldots h_k}$ of $B$
	and every index $1\leq i\leq k$ such that $h$ and $h_i$ belong to the same strongly connected component $G_B$,
	it holds that $h_i=\top$ for all $i\neq j$.
\end{itemize}
\end{lemma}

%
\begin{proof}
Assume that for every transition $\trans{h,f,h_1\ldots h_k}$ of $B$,
$h_j\neq\top$ for at most one index $j\in\{1,\ldots,k\}$.
Then we construct a \DBT\ $A$ with $H$ as set of states by introducing for for every transition $\trans{h,f,h_1\ldots h_k}$ of $B$,
one rule. 
If there is no $j$ with $h_j\neq\top$, the rule takes the form 
\[
h(f(x_1,\ldots,x_k)) \to a
\]
and if $h_j\neq\top$ for some $1\leq j\leq k$, the rule is given by
\[
h(f(x_1,\ldots,x_k)) \to h_j(x_j)
\]
It is easily checked that the domain of this transducer is precisely the language accepted by $B$.
For the reverse direction, assume that $B$ has a transition $\trans{h,f,h_1\ldots h_k}$ such that $h_{j_1},h_{j_2}$ are distinct from $\top$
for indices $1\leq j_1 < j_2\leq k$. 
Since $B$ is trim, there is a context $c$ and input trees $t_1,\ldots,t_k$ and $t'_{1},t'_{2}$ such that 
\begin{itemize}
\item	$(h_0,h): c$ for the initial state $h_0$ of $B$; 
\item	$t'_{1}\not\in\dom_B(h_{j_1})$ and $t'_{2}\not\in\dom_B(h_{j_2})$; but
\item	$h_i:t_i$ for all $i=1,\ldots,k$.
\end{itemize}
Accordingly, $u = c\cdot f(t_1,\ldots,t_k)\in\L(B)$ holds, but both
$u_1 = c\cdot f(t_1,\ldots,t_{j_1-1},t'_{1},t_{j_1+1},\ldots,t_k)$ and
$u_2 = c\cdot f(t_1,\ldots,t_{j_2-1},t'_{2},t_{j_2+1},\ldots,t_k)$ are not.
For a contradiction assume that there is \DBT\ $A$ whose domain coincides with $\L(B)$ and axiom $s$.
If $s$ is ground, then the domain of $A$ equals all $\T_\Sigma$, and thus is different from $\L(B)$.
Therefore, assume that $s$ is not ground, but of the form $s'\cdot q_0(x_1)$.
Since the domain of $A$ equals $\L(B)$, there is a computation $(q_0,h):c\to v$ of 
$A$ on context $c$.
If $v$ is ground, the domain of $A$ contains not only $u$, but also $u_1$ and $u_2$ -- in contradiction to our assumption.
So assume that $v = v'\cdot q'(x_1)$ for some $v'\in\T_\Delta(x_1)$ and state $q'$ with $\iota(q')=h$.
Since $A$ is defined for $u$, there must be some transition $q'(f(x_1,\ldots,x_k))\to T$ of $A$ for $q'$ and $f$.
Since the every $a\in\Delta$ has rank at most 1, the right-hand side $T$ has at most one occurrence of a term $q_i(x_i)$.
Thus, $T$ cannot have occurrences both of a variable $x_{j_1}$ and a variable $x_{j_2}$ --
implying that $A$ is defined for at least one tree in $\{u_1,u_2\}$ -- contradiction.

Now consider the second statement of the lemma.
First assume that for every transition $\trans{h,a,h_1\ldots h_k}$ of $B$
and every index $1\leq i\leq k$ such that $h$ and $h_i$ belong to the same strongly connected component $G_B$,
it holds that $h_i=\top$ for all $i\neq j$. Then we construct a \DBT $A$ with domain $\L(B)$ as follows.
Let $n$ denote the number of non-trivial strongly connected components, 
Let $p\in\T_\Delta(X_r)$
where $r$ is the maximal rank of an input symbol $f\in\Sigma$. Since $\Delta$ has some element of rank at least 2,
such a pattern $p$ exists. Let $a\in\Delta$ denote some output letter of rank 0. 
Then we define $T_0 = a$ and for $n>0$, $T_n = p\{x_j\mapsto T_{n-1}\mid j = 1,\ldots,r\}$.
Let $N$ denote the number of strongly connected components of $G_B$.

Then we define a \DBT\ $A$ as follows the set of states of $A$ are pairs $\angl{h,n}$ where $h$ is a state
of $B$ and $n\in\{0,\ldots,N\}$ where the axiom is given by $\angl{h_0,0}$.
Intuitively, the first component of a state records the current state of the deterministic top-down TA $B$,
and the second component records the number of strongly connected components of $G_B$ which have already been traversed.
Now consider a transition $\trans{h,f,h_1\ldots h_k}$ of $B$ and $\angl{h,n}$ is a state attained by $A$.
\begin{itemize}
\item	If there is some $j\in\{1,\ldots,k\}$ such that $h$ and $h_j$ belong to the same strongly connected component of $G_B$,
	then $A$ has the transition
	\[
	\angl{h,n}(f(x_1,\ldots,x_k))\to \angl{h_j,n}(x_j)
	\]
\item	Now assume that this is not the case. 
	For $i=1,\ldots,r$, let $s_i = \angl{h_i,n+1}(x_i)$ whenever $i\leq k$ and $h_i\neq\top$,
	$s_i = T_{N-n-1}$ otherwise. Then $A$ has the transition
	\[
	\angl{h,n}(f(x_1,\ldots,x_k))\to p\{x_i\mapsto s_i\mid i=1,\ldots,N\}
	\]
\end{itemize}
Then the domain of $A$ equals $\L(B)$ where the output for every $t\in\L(B)$ equals $T_N$.

For the reverse direction, assume that there is a transition $\trans{h,f,h_1\ldots h_k}$
where there are indices $i\neq j$ so that $h$ and $h_i$ are in the same strongly connected component of $G_B$,
but $h_j$ is different from $\top$.
W.l.o.g., assume that $i < j$.
Since $h_j\neq\top$, there is some input tree $t'\not\in\dom(h_j)$.
Since $B$ is trim and $h$ and $h_i$ belong to the same strongly connected component of $G_B$,
there are contexts $c_c'$ and input trees $s_1,\ldots,s_k$ together with computations
$(h_0,h):c,(h_1,h):c_1$ and $h_\nu:s_\nu$ for all $\nu=1,\ldots,k$. In particular
for $c_f = f(s_1,\ldots,s_{i-1},c_1\cdot x_1,s_{i+1},\ldots,s_k)$, 
$(h,h):c_\ell$ holds.
Moreover, $u_0 = c\cdot f(s_1,\ldots,s_k)\in\L(B)$ as well as
$u_\ell = c\cdot c_f^\ell\cdot s_i\in\L(B)$ for all $\ell\geq 0$.
Let 
\[
c'_f = f(s_1,\ldots,s_{i-1},c_1\cdot x_1,s_{i+1},\ldots,s_{j-1},s'_j,s_{j+1},\ldots,s_k)
\]
Then for every $k_1,k_2\geq 0$, the tree
$u'_{k_1,k_2} = c\cdot c_f^{k_1}\cdot c'_f \cdot c_f^{k_2}\cdot s_i\not\in\L(B)$.
Now assume that $A$ is a \DBT~ with domain $\L(B)$ which maps all input trees $t\in\L(B)$ to the same output tree $T$.
As in the case where all arities of output symbols is bounded by 1,
the axiom of $A$ must be of the form $s'\cdot q_0(x_1)$ for some state $q_0$ of $A$.
Let $(q_0,h):c\to v$. Then $v$ must be necessarily of the form $v'\cdot q_1(x_1)$ (otherwise, $A$ would not only be defined
for $u_0$, but also $u'_{k_1,k_2}$ for any $k_1,k_2\geq 0$).
Likewise for every $\ell\geq 0$,
$(q_1,h):c_f^\ell\mapsto v'_\ell\cdot q_\ell(x_1)$ for some $v'_\ell\in\T_\Delta(x_1)$ and state $q_\ell$ of $A$.
Consequently, there must be exponents $\ell_1,\ell_3\geq 0$, $\ell_2>0$ and some state $q$ such that
$(q_0,h_0): c\cdot c_f^{\ell_1} \to v_1\cdot q(x_1)$, 
$(q,h): c_f^{\ell_2} \to v_2\cdot q(x_1)$, and
$q: c_f^{\ell_3}\cdot s_i \to v_3$ holds.
This means that for every $m\geq 0$, the output tree $s'\cdot v_1\cdot v_2^m\cdot v_3$ can be produced. 
Since all these trees must equal the single tree $T$,
we conclude that $v_2$ must equal $x_1$. But then we can replace $s_j$ in the first factor $c_f$ of $c_f^\ell$ with
$s'_j$, i.e., consider the tree
\[
u'= c\cdot c_f^{\ell_1}\cdot c'_f\cdot c_f^{\ell_2 -1+\ell_3}\cdot s_i
\]
and find that $A$ produces $T$ for $u'$ as well. Since $u'\not\in\L(B)$, we have a contradiction.
\end{proof}
}
Then we have:
\begin{lemma}\label{l:output-recognizable}
Let $s$ denote some output tree in $\T_\Delta$ and $h\neq\top$ a state of $B$.
Then $\dom_B(h)$ is output-recognizable by $s$ iff for every transition 
$\tau = \trans{h,f,h_1\ldots h_k}$ with $J = \{j\in\{1,\ldots,k\}\mid h_j\neq\top\}$
there is a pattern $T_{s,\tau}\in\T_\Delta(X_J)$ such that
$s$ can be decomposed into
$s = T_{s,\tau}\{x_j\mapsto s_j\mid j\in J\}$ so that $\dom_B(h_{j})$ is output-recognizable by $s_j$ 
for all $j\in J$.
\ignore{
In case that $h$ and one of the $h_j$ are contained in the same strongly connected component of $G_B$,
then $r$ must equal 1, and $s_\tau$ can be chosen as $x_1$.
}
\end{lemma}
\begin{proof}
Using the patterns $T_{s',\tau'}$ for subtrees $s'$ of of $s$, we can construct a 
\DBT\ $A_{s}$ with states $\angl{s',h'}$ for
subtrees $s'$ of $s$ and states $h'$ of $B$, 
an axiom $\alpha = \{h\mapsto\angl{s,h}\}$, 
and rules
\[
\angl{s',h'}(f(x_1,\ldots,x_k))\to T_{s',\tau'}
\{x_j\mapsto \angl{s_j,h_{i_j}}(x_{i_j})\mid j\in J\}
\]
for every transition $\tau' = \trans{h',f,h_1\ldots h_k}$ with 
$J= \{j\in\{1,\ldots,k\}\mid h_j\neq\top\}$
and trees $s_j$ such that $s' = T_{s',\tau'} \{x_j\mapsto s_j\mid j\in J\}$.
Then the domain of $A_{s}$ exactly equals $\dom_B(h)$ where every input tree in $\dom_B(h)$
is mapped to $s$.

For the reverse direction, 
assume that $h\in H$ is output-recognizable by $s$, i.e., there is a transducer
$A$ without inspection 
such that the domain of $A$ equals $\dom_B(h)$ and
$\sem{t}_{A} = s$ for all $t\in\dom_B(h)$.
Assume that the axiom of $A$ is given by $\{h\mapsto T_0\cdot q(x_1)\}$. Then
$s = T_0\cdot s_0$ where $\sem{t}_{A,q} =s_0$ for all $t\in\dom_B(h)$.
Moreover for every other reachable state $q'$ of $A$ with $\iota(q') = h'$, 
we have for some subtree $s_{q'}$ of $s_0$ such that 
$\sem{t}_{A,q'} = s_{q'}$ holds for all $t\in\dom_B(h)$.
W.l.o.g., we may assume that the prefix $T_0$ of the axiom is \emph{empty}, i.e., equals $x_1$:
If this is not yet the case, we can introduce a new state $q_0$ with $\iota(q_0) = h$
together with rules
\[
q_0(f(x_1:h_1,\ldots,x_k:h_k)) \to T_0\cdot T
\]
whenever $A$ has a rule 
\[
q(f(x_1:h_1,\ldots,x_k:h_k)) \to T
\]
and replace the axiom of $A$ with $\{h\mapsto q_0(x_1)\}$.
Assume now that $A$ is such a transducer with a \emph{minimal} number of states.
Two further properties of $A$ can then be derived:
\begin{itemize}
\item	For no state $q'$ of $A$, $\iota(q')=\top$, since since every occurrence of $q'(x_i)$ 
	in a right-hand side can be equivalently replaced with $s_{q'}$ -- thus effectively removing $q'$ 
	from $A$.
\item	For all $q\neq q'$ with $\iota(q) = \iota(q')$, the corresponding output trees
	$s_{q}$ and $s_{q'}$ must differ. If they were equal and, say, $q'$ is different from $q_0$,
	then all occurrences of $q'(x_i)$ in right-hand sides could be replaced with $q(x_i)$ -- 
	thus eliminating $q'$ from the transducer.
\end{itemize}
Now consider a transition $\tau =\trans{h',f,h_1\ldots h_k}$ of $B$ where 
$J = \{j\in\{1,\ldots,k\}\mid h_j\neq\top\}$.
Then there must be some $q'$ with $\iota(q') = h'$ so that $A$ has a rule
\[
q'(f(x_1:h_1,\ldots,x_k:h_k)) \to T
\] 
where $T$ is of the form $T = s'\{x_j\mapsto q_j(x_j)\mid j\in J\}$.
Then we can define the pattern $T_{s_{q'},\tau}$ as $s'$.
The collection of these patterns $T_{s_{q'},\tau}$ satisfy the assumptions of the lemma.
\ignore{
First assume that $h'$ and some $h_j$ are in the same strongly connected component. 
We claim that then $T = q'(x_j)$ for some state $q'$ with $\iota(q') = h_j$:
Assume for a moment that this were not the case. Then arbitrary large outputs can be
produced by $A_s$ for input trees in $\dom_B(h')$ and thus also from $\dom_B(h)$.
}
\end{proof}
From the construction in the proof of Lemma \ref{l:output-recognizable} we further conclude that 
the patterns $T_{s,\tau}$ can in fact be chosen \emph{linear}, i.e., each 
variable $x_j$ occurs exactly once. In this case, the constructed transducer is a \DLT.
According to Lemma \ref{l:output-recognizable},
the set of all states $h$ such that $\dom_B(h)$ is output-recognizable by $s$, 
can be determined by tabulating for each state $h'$ of $B$ the set of \emph{minimal} subtrees $s'$ of $s$
so that $\dom_B(h')$ is output-recognizable by $s'$. Note that the minimal subtrees suffice since 
$\dom_B(h)$
is output-recognizable for $s$ already whenever $\dom_B(h)$ is output-recognizable for some subtree 
of $s$.

\begin{example}\label{e:binary-1}
The language $L$ from Example \ref{e:binary} can be accepted by the top-down automaton $B$ with set if states
$H = \{h_0,h_a,h_b,\top\}$ and transitions
\[
\trans{h_0,f,h_a h_b},
\trans{h_a,f,h_a \top},
\trans{h_a,a},
\trans{h_b,f,\top h_b},
\trans{h_b,b}.
\]
Then the set of minimal subtrees of $g(c,c)$ for which $\dom_B(h_a)$ and $\dom_B(h_b)$ 
are output-recognizable is $\{c\}$,
while $\dom_B(h_0)$ is output-recognizable only by $g(c,c)$.
\qed
\end{example}

\noindent
In summary, we obtain:

\begin{theorem}\label{t:output}
Assume that the input alphabet is fixed.
%
%
Then for a given state $h$ of a top-down deterministic TA $B$, 
	it can be decided in polynomial time whether or not
	$\dom_B(h)$ is output-recognizable by a particular tree $s$ and if so,
	a \DLT\ $A_{B,h,s}$ without inspection can be constructed in polynomial time with domain
	$\dom_B(h)$ such that $\sem{t}_{A_{B,h,s}} = s$ for all $t\in\dom_B(h)$.
	\qed
\end{theorem}

%
%

\section{How to Satisfy Inspection Needs}\label{s:need}

In the following, we consider an arbitrary \DBTI transducer $A$
with underlying top-down deterministic TA $B$
as \emph{inspection} automaton. We recall that,
since $B$ has only a single accepting state, $A$ only provides a single axiom.
Consider a rule $\tau$ of the form $q(f(x_1:h_1,\ldots,x_k:h_k)) \to T$ of $A$.
For every $x_i$ not occurring in $T$, it must be verified that
the corresponding subtree of the input is contained in $\dom_B(h_i)$.
This verification is not needed if $h_i=\top$, i.e., $\dom_B(h_i)=\T_\Sigma$.
%
Accordingly, let $J_\tau$ denote the set of indices $j$ such that $x_j$ does \emph{not}
occur in $T$ while at the same time, $h_{j}\neq \top$. 
Let us thus call the set
$\eta_\tau = \{(j,h_{j})\mid j\in J_\tau\}$ the \emph{inspection need} of the rule $\tau$.

Assume that $T$ has disjoint ground subtrees $s_j,j\in J_\tau$,
such that $\dom_B(h_{j})$ is output-recognizable by $s_j$ for $j\in J_\tau$.
Then the rule $\tau$ can equivalently be replaced by rules without inspection need.
This can be seen as follows.
For each $j\in J_\tau$, consider the transducer $A_j = A_{B,h_j,s_j}$ from Theorem \ref{t:output} 
with domain $\dom_B(h_j)$ and single output $s_j$. By definition, each state of $A_j$ only produces a single output.
W.l.o.g., assume that the sets of states of the $A_j$ are all disjoint and different from the states of $A$.
Then $\tau$ is replaced with the rule obtained from $\tau$ by replacing each $s_j$ in $T$
with the tree obtained from the axiom of $A_j$ by substituting $x_1$ with $x_j$; 
furthermore, all rules from the $A_j$ are included.
Therefore in this case, we say that $\tau$ 
\emph{satisfies} its inspection need.
\begin{example}\label{e:need 2}
	Consider the rule $\tau$ given by
	\[
	q(f(x_1:h_1,x_2:h, x_3:h))\to g(q'(x_1), r(b), b)
	\]
	where $\dom_B(h)$ equals the set $L=\{ g(a,t)\mid t\in\T_\Sigma\}$.
	Then $J_\tau=\{2,3\}$, and
	the inspection need of the rule $\tau$ is $\eta_\tau=\{(2,h),(3,h)\}$ because we need to verify for the 
	input subtrees corresponding to the variables $x_2$ and $x_3$ in $\tau$
	whether or not they are contained in $L=\text{dom}_B (h)$.
	The language $L$ is output-realizable by $r(b)$ as well as $b$. 
	One possible way of removing inspection therefore is by the following rules:
	\[
	\begin{array}{lll}
	q(f(x_1,x_2, x_3))	&\to& g(q'(x_1), r(q_1(x_2)), q_1(x_3))	\\
	q_1(g(x_1,x_2))		&\to& q_2(x_1)	\\
	q_2(a)			&\to& b.
	\end{array}
	\]
	\qed
\end{example}

\noindent
Our goal is to construct for a given \DBTI transducer $A$ an equivalent \DBTI trans\-ducer $A'$
such that each rule of $A'$ satisfies its inspection need.
If each rule of $A$ satisfies its inspection need, 
this need no longer be the case for the \emph{earliest} transducer equivalent to $A$.
The reason is that
some ground subtrees of prefixes of right-hand sides may have been moved to the right-hand sides
of other rules.
Satisfying inspection needs of rules therefore requires to partly 
\emph{revert} the \emph{earliest} transformation.
\ignore{
In the following, we call a state $q$ of the \DBTI\ $A$ \emph{constant},
if $q$ can only produce a single output tree $s$, i.e., for all trees $t$ in the domain of $q$, $q:t\to s$ holds.

\begin{lemma}\label{l:const}
For a partial mapping $\mu:\T_\Sigma\to\T_\Delta$, the following 
are equivalent:
\begin{enumerate}
\item	
$\mu$ is realized by a \DBT\ without inspection;
\item	
$\mu$ is realized by a \DBTI\ without constant states, but
 	where all inspection needs are satisfied.
	\qed
\end{enumerate}
\end{lemma}
}

\subsection{Analyzing Inspection Needs}\label{ss:needs}

\newcommand{\semsharp}[1]{{\sem{ #1 }^\sharp}}
\newcommand{\semsharpsharp}[1]{\sem{ #1 }^{\sharp\sharp}}

Let $A$ denote a \DBTI\ in canonical earliest normal form, and assume that $A$ is equivalent to some \DBT\ $A'$.
Recall that each state of $A$ can be considered as an equivalence class of non-constant states of $A'$,
where for each state $q$ of $A$ and each state $q'\in q$ of $A'$ with 
$\pref_{A'}(q') = p$, it holds that 
$p \cdot \sem{t}_{A,q} = \sem{t}_{A',q'}$. 
In particular, the constant outputs for some states of $A'$ may 
occur as subtrees in $p$ and thus are already produced before $A$ 
processes the input $t$.

In order to recover the (yet unknown) \DBT\ $A'$ without inspection from the canonical earliest \DBTI\ $A$, 
we determine the minimal suffix $p'$ of $p$ so that all inspections possibly 
encountered when $q$ processes its input, can be satisfied.

\begin{example}\label{e:generalized}
    Let $h_a$, $h_b$ be inspection states.
	Let the domains of $h_a$, $h_b$ consist of all trees whose left-most leaf is labeled by $a$
	or $b$, respectively.
	Consider the canonical earliest \DBTI $A$ with axiom $\alpha(h_0) = f(a,f(a,q_0(x_1)))$ and the rules
	\[
	\begin{array}{lll}
	q_0 (f (x_1: h, x_2:h_a)) 	&\to & q_1(x_1)  \\
	q_1 (f (x_1: h, x_2:h_b)) &\to& q_2 (x_1)\\
 	q_1 (g (x_1: h, x_2:h_b, x_3: h_b)) &\to & f(a, q_2 (x_1)) \\
 	q_2 (a)  &\to& a\,.
	\end{array}
	\]
	%
	No rule for non-leaf input symbols satisfies its inspection need.
	We may, however, delay the output of the axiom in such a way that the right-hand sides of these rules receive 
	$f(a,x_1)$ as an additional prefix. For the axiom $\alpha'(h_0) = q_0(x_1)$, we thus obtain the new set of rules
	\[
	\begin{array}{lll}
	q_0 (f (x_1: h, x_2:h_a)) 	&\to & f(a,q_1(x_1))  \\
	q_1 (f (x_1: h, x_2:h_b)) &\to& f(a, q_2 (x_1))\\
 	q_1 (g (x_1: h, x_2:h_b, x_3: h_b)) &\to & f(a, f(a, q_2 (x_1))) \\
 	q_2 (a)  &\to& a\,.
	\end{array}
	\]
	Now the inspection needs of all rules are satisfied. Hence, an equivalent transducer without inspection can
	be constructed.
	\qed
\end{example}

\noindent
Example~\ref{e:generalized} demonstrates that ground suffixes of an axiom  $p\{x_1 \mapsto  q_0(x_1)\}$
with $p\in T_\Delta (x_1)$ may need to satisfy inspection needs that arise when processing input trees starting in
state $q_0$.
%
%
Likewise for a rule $\tau$ of the form $q (f (x_1:h_1,,\dots, x_k:h_k))\rightarrow p\{x_j\mapsto  q_j(x_j)\mid j\in J\}$
with $p\in T_\Delta (X_J)$,
it may not be sufficient only to satisfy the inspection needs of $\eta_\tau$; for all $j\in J$,
ground suffixes of $p$ may further be required to satisfy inspection needs that subsequently arise when subtrees are to be
processed starting from the $q_j$.
Such a \emph{generalized inspection need} of an input tree to be processed by some state $q$
is represented by a sequence 
\[
(M_1,\{\emptyset\})\ldots(M_{r-1},\{\emptyset\})(M_r,\phi)
\]
where $r\geq 0$, $M_1,\ldots,M_r$ are sets of inspection needs each corresponding to one rule of the transducer, and
$\phi$ is a \emph{downward-closed subset} of subsets of $M_r$, i.e.,
if the subset $M$ of $M_r$ is contained in $\phi$ then
every  subset of $M$ is also contained in $\phi$.
Intuitively, a generalized inspection need is the \emph{sequence} of future inspection needs yet to be 
satisfied when the appropriate node in the input tree is encountered. 
The pair $(M_r,\phi)$ is meant to occur farest in the future.
Thereby, $\phi$ is the set of all subsets of individual inspections of $M_r$ that can 
already be accomplished (the available ground subtrees might be used in different ways).
%
%
Thus, if $\phi=\{\emptyset\}$, then none of the individual inspections can be accomplished,
and if $M_r$ itself is contained in $\phi$, then \emph{all} languages 
$\dom_B(h), (j,h)\in M_r,$ 
are simultaneously realizable --- 
in which case we drop the pair $(M_r,\phi)$ as a whole.
Therefore, we generally assume that $M_r\not\in\phi$.
%
%
\begin{example}\label{e:generalized-1}
	Consider the \DBTI $A$ of Example \ref{e:generalized} and let $\tau_i,i=1,2,3,4$, denote $i$th rule of $A$.
	Then the sets $\eta(\tau_i)$ are given by
	\[
	\begin{array}{|c|l|}	
	\hline
	i	&	\eta_{\tau_i}	\\
	\hline\hline
	1	&	\{(2,h_a)\}	\\
	2	&	\{(2,h_b)\}	\\
	3	&	\{(2,h_b), (3,h_b)\}	\\
	4	&	\emptyset	\\
	\hline
	\end{array}
	\]
	Trees in the domain of $q_0$ are either of the form
	\[
	 f(f (a, s_2), s_1)\quad\text{or}\quad 	f(g (a, s_2, s_3), s_1)
	\]
	where the left-most leaf of $s_1$ is labeled with $a$
	and the left-most leaves of $s_2$ and $s_3$ are labeled with $b$.
	%
	%
	In the former case, the generalized inspection need is 
	\[
	(\eta_{\tau_1},\{\emptyset\}) (\eta_{\tau_2},\{\emptyset\}) = (\{(2,h_a)\},\{\emptyset\})\,(\{(2,h_b)\},\{\emptyset\})
	\]
	In the latter case, the generalized inspection need is 
	\[
	\begin{array}{rll}
	(\eta_{\tau_1},\{\emptyset\}) (\eta_{\tau_3}, \phi) &=& (\{(2,h_a)\},\{\emptyset\})\,
	   (\{(2,h_b),(3,h_b)\},\phi) \qquad\text{where}	\\
	   \phi	&=&\{\{(2,h_b)\},\{(3,h_b)\},\emptyset\}
        \end{array}
\]
	where $\phi$ indicates that
	the ground subtree $a$ of $\tau_3$ can be used to check one of the
	inspection needs $(2,h_b)$ or $(3,h_b)$ from $\eta_{\tau_3}$ or none.
	\qed
\end{example}

\noindent
Subsequently, we show how all possible generalized inspection needs for states $q$ can be computed.
Let $\In$ denote the set of all possible generalized inspection needs.

For a finite multiset $G$ of trees in $\T_\Delta$ and a finite set $M\subseteq\mathbb{N}\times H$,
define $\angl{\angl{G,M}}$ as the set of all subsets $M'$ of $M$ so that 
the multiset of languages $\{\dom_B(h)\mid \exists\,j\,\text{such that}\,(j,h)\in M'\}$ is simultaneously output-recognizable
by means of disjoint subtrees of trees in $G$.
The set $\angl{\angl{G,M}}$ is downward closed, i.e.,
if the multiset $\{\dom_B(h)\mid \exists\,j\,\text{such that}\,(j,h)\in M'\}$ is simultaneously output-recognizable
by means of disjoint subtrees of trees in $G$,
then this also holds for every subset $M''$ of $M'$.
Technically, the set $\angl{\angl{G,M}}$ thus may be represented by its maximal elements.

\begin{example}\label{e:example}
	Let $\tau$ be the rule 
	$q(f(x_1:h_1,x_2:h_2)) \to a$
	with inspection need $\eta_\tau=\{(1,h_1),(2,h_2)\}$.
	Assuming that $\dom_B(h_i)$ is output-recognizable by $a$,
	only one of the inspection needs of the rule $\tau$ can be satisfied by $\tau$.
	Therefore,  $\angl{\angl{\{a\},\eta_\tau}}=\{ \{(1,h_1)\}, \{(2,h_2)\},\emptyset \}$
	with maximal elements $\{(1,h_1)\},\{(2,h_2)\}$.
	If, however, the right-hand side of $\tau$ had been $f(a,a)$,
	then all inspection needs of $\tau$ would be satisfiable.
	In this case, \[\angl{\angl{\{f(a,a)\},\eta_\tau}} =
	\{ \{(1,h_1),(2,h_2)\}, \{(1,h_1)\},\{(2,h_2)\},\emptyset \}\] with single maximal element 
	$\{(1,h_1),(2,h_2)\}$.
\qed
\end{example}

\noindent
Now assume that $p\in\T_\Delta(x_1)$ where we write $\G[p]$ for the multiset of ground subtrees
occurring in $p$.
\ignore{
$p = s_1\cdot \ldots\cdot s_m$ where each of the $s_j\in\T_\Delta(x_1)$ is
\emph{irreducible}, i.e., cannot be written as a product of patterns each different from $x_1$. 
For convenience, we then define the \emph{length} $|p|$ of $p$ as $m$.
}
For $p$, we define the auxiliary transformation $\semsharpsharp{p}:\In\to\In$ on generalized inspection needs.
This transformation calculates for a given generalized inspection need $\alpha$ the remaining generalized inspection need
which cannot be satisfied by the ground subtrees occurring in $p$.
We proceed by induction on the length of $\alpha$.
\ignore{
It is obtained as the composition
$\semsharpsharp{s_1}\circ\ldots\circ\semsharpaharp{s_m}$.
where the ground subtrees of each $s_j$ can only be used to satisfy the inspection needs of a single rule, i.e.,
one factor of the generalized inspection need.
For an irreducible pattern $s\in\T_\Delta(x_1)$, the transformation $\semsharpsharp{s}\alpha$ thus is
defined by as follows. 
}
\begin{itemize}
\item	If $\alpha=\epsilon$, i.e., is empty, then 
	$\semsharpsharp{p}\epsilon = \epsilon$. 
\item	Let $\alpha = \alpha' (M,\phi)$ for some set $M$. 
	For each suffix $v$ of $p$, we calculate the set
	\[
	\phi[v] = \{R_1\cup R_2\mid R_1\in\phi,R_2\in\angl{\angl{\G[v],M\setminus R_1}}\}.
	\]
	Then $v$ suffices to satisfy all inspection needs of $M$ left open by $\phi$
	iff $M\in\phi[v]$.

	First assume that there is no such $v$. Then
	$\semsharpsharp{p}\alpha = \alpha'(M,\phi[p])$.

	Otherwise, let $p=u\cdot v$ where $v$ is the minimal suffix with
	$M\in\phi[v]$. Then 
	$\semsharpsharp{p}\alpha = \semsharpsharp{u}\alpha'$.
\end{itemize}
	%
%
%
For an input tree $t\in\T_\Sigma$, we define
the inspection need $\semsharp{t}_{A,q}$ of a state $q$ of $A$ for $t$ as follows.
Assume that $t=f(t_1,\ldots,t_k)$ and $\tau$ is the rule of $A$ of the form
$q(f(x_1:h_1,\ldots,x_k:h_k)) \to p\{x_j\mapsto q_j(x_j)\mid j\in J\}$ 
such that $p\in\T_\Delta(X_J)$ holds with $h_i:t_i$ for $i=1,\ldots,k$. 
For the rule $\tau$, we define a transformation
\[
\semsharp{\tau}:(X_J\to\In)\to\In
\text{ such that }
\semsharp{t}_{A,q} = \semsharp{\tau}\{x_j\mapsto\semsharp{t_j}_{A,q_j}\mid j\in J\}
\text{ holds.}
\]
The transformation $\semsharp{\tau}$ is defined by case distinction.
\begin{itemize}
\ignore{
\item
If $J= \emptyset$, i.e., $p$ is ground, we check in how far $p$ itself is sufficient
for $\eta_\tau$ to be output-realizable. Let $\phi = \angl{\angl{\{p\},\eta_\tau}}$.
Then
\[
\semsharp{\tau}() = \left\{
	\begin{array}{l@{\quad}l}
	\epsilon	&\text{if}\;\eta_\tau\in\phi	\\
	(\eta_\tau,\phi)&\text{otherwise.}
	\end{array}\right.
\]
}
\item
If $J=\{j\}$, then  
\[
\semsharp{\tau}\{x_j\mapsto\alpha\} = \semsharpsharp{p}((\eta_\tau,\{\emptyset\})\,\alpha).
\]
\item
Otherwise, $J$ is empty or contains more than one index.
Let $p=p'\{x_j\mapsto p_j\cdot q_j(x_j)\mid j\in J\}$ for
\emph{maximal} patterns $p_j\in\T_\Delta(x_1)$.
For $\alpha_j,j\in J$, assume that $p_j = p'_j\cdot u_j$ for some
\emph{minimal} suffix $u_j\in\T_\Delta(x_1)$ with $\semsharpsharp{u_j}\alpha_j = \epsilon$.
These suffixes $u_j$ must exist, whenever $A$ is equivalent to some \DBT\ without inspection
and the generalized inspection needs $\alpha_j$ originate from some input tree from $\dom_B(\iota(q_j))$.
Let $G$ denote the multiset of distinct maximal ground subtrees of 
$p'\{x_j\mapsto p'_j\cdot x_j\mid j\in J\}$, and $\phi=\angl{\angl{G,\eta_\tau}}$.
Then we define
\[
\semsharp{\tau}\{x_j\mapsto\alpha_j\mid x_j\in X_J\} = \left\{
	\begin{array}{ll}
	\epsilon			&\quad\text{if}\;\eta_\tau\in\phi	\\
	(\eta_\tau,\phi)		&\quad\text{otherwise.}
	\end{array}\right.
\]
\end{itemize}

\begin{example}\label{e:generalized-2}
	Consider again the \DBTI $A$ from Example~\ref{e:generalized}.
        The most obvious transformation is $\semsharp{\tau_4}$. 
	The right-hand side of $\tau_4$ equals the ground tree $a$,
	and there is no inspection need. Therefore,
	\[
	\semsharp{\tau_4} = \epsilon\,.
	\]
	For $i=1,2,3$, the transformations $\semsharp{\tau_i}$ are given by:
	\[
	\begin{array}{|c|l|l|}	
	\hline
	i&\text{rhs}	&	\semsharp{\tau_i}\,\alpha	\\
	\hline\hline
	1&q_1(x_1)		&	(\{(2,h_a)\},\{\emptyset\})\;\alpha	\\
	2&q_2(x_1)		&	(\{(2,h_b)\},\{\emptyset\})\;\alpha	\\
	3&f(a,q_2(x_1))	&	\semsharpsharp{f(a,x_1)}((\{(2,h_b), (3,h_b)\},\{\emptyset\})\;\alpha)	\\
	\hline
	\end{array}
	\]
	For ease of understanding, we have again listed the right-hand sides of $\tau_i$ in the
	second column. 
	In all three cases, the set of indices of occurring variables equals $\{1\}$.
	Another inspection need is added to $\alpha$. The first two rules do not provide
	any further ground subtrees, while the third rule has the factor $f(a,x_1)$ which can be used
	to satisfy the accumulated generalized inspection need.
	
	These transformations, we can now use to determine the set of all generalized inspection needs of $q_0$ 
	for input trees accepted by $A$.
	Recall that these input trees 
	are either of the form $f(f (a, s_2), s_1)$ or $f(g (a, s_2, s_3), s_1)$
	where the left-most leaf of $s_1$ is labeled with $a$
	and the left-most leaves of $s_2$ and $s_3$ are labeled with~$b$.
	
	First consider input trees of the form $s = f(f (a, s_2), s_1)$.
	We have:
	\[
	\begin{array}{lll}
	\eta_{q_0}(s) 	
			&=& \semsharp{\tau_1} (\semsharp{\tau_2}(\semsharp{\tau_4}))	\\
			&=& \semsharp{\tau_1} (\semsharp{\tau_2}(\epsilon))	\\
			&=& \semsharp{\tau_1} (\{(2,h_b),\},\{\emptyset\})	\\
			&=& (\{(2,h_a)\},\{\emptyset\})\; (\{(2,h_b)\},\{\emptyset\}).
	\end{array}
	\]
Now consider input trees $s' = f(g(a,s_2,s_3), s_1)$
where again the left-most leaf of $s_1$ is labeled with $a$
and the left-most leaves of $s_2$ and $s_3$ are labeled with~$b$.
The generalized inspection need of $s'$ then is given by 
\[
\begin{array}{lll}
\eta_{q_0} (s') 
			&=& \semsharp{\tau_1} (\semsharp{\tau_3}(\semsharp{\tau_4}))	\\
			&=& \semsharp{\tau_1} (\semsharp{\tau_3}(\epsilon))	\\
			&=& \semsharp{\tau_1} (
				\semsharpsharp{f(a,x_1)}(
				(\{(2,h_b),(3,h_b)\},\{\emptyset\})))	\\
			&=& \semsharp{\tau_1}(
				(\{(2,h_b),(3,h_b)\},\phi))	\\
			&=& (\{(2,h_a)\},\{\emptyset\})\; (\{(2,h_b),(3,h_b)\},\phi)	\\
	\end{array}
\]
where $\phi = \{ \{(2,h_b)\},\{(3,h_b)\},\emptyset\}$.
\qed
\end{example}

\noindent
For convenience, we extend the transformation $\semsharp{\cdot}$ to contexts $t\in\C_\Sigma$ by
\[
\begin{array}{lll}
\semsharp{x_1}_{A,q}^h (\alpha)	&=&	\alpha		\\
\semsharp{f(t_1,\ldots,t_k)}_{A,q}^h (\alpha)	&=&	\semsharp{\tau}\{x_j\mapsto \semsharp{t_j}_{A,q_j}^h,
	x_{j'}\mapsto\eta_{q_{j'}}(t_{j'})\mid j'\in J\setminus\{j\}\}
\end{array}
\]
where in the first case, $\iota(q) = h$,
and in the second, $\tau$ is the rule
$q(f(\ldots))\to p\{x_{j'}\mapsto q_{j'}(x_{j'})\mid {j'\in J} \}$,
$t_j\in\C_\Sigma$ with $j\in J$, and $t_1,\ldots,t_{j-1},t_{j+1},\ldots,t_k\in\T_\Sigma$.
We have:

\begin{lemma}\label{l:correctness}
Assume that $A$ is the canonical earliest normal form of some \DBTI $A'$ without inspection needs.
Let $q$ denote some state of $A$, i.e., an equivalence class of states of $A'$.
Let $q'\in q$ be state of $A'$, let $p=\pref_{A'}(q')$ the maximal common prefix of outputs
of $A'$ for $q'$, and $t\in\dom_B(\iota(q))$.
Then 
\begin{enumerate}
\item	$\semsharp{t}_{A,q}$ is defined; and
\item	$\semsharpsharp{p}(\semsharp{t}_{A,q}) = \epsilon$.
\end{enumerate}
\qed
\end{lemma}
\noindent
The proof is by induction on the structure of $t$ where we use that 
$\sem{t}_{A',q'} = p\cdot \sem{t}_{A,q}$ holds for $p=\pref_{A'}(q')$ .


By computing $\semsharp{t}_{A,q}$, we partly recover information on the (unknown) common prefix $p$
of the (yet to be constructed) \DBTI $A'$ for $q'$.
%
%
By computing the set of \emph{all} generalized inspection needs for $q$, 
we determine the maximal requirement on a suffix of output already produced by $A$ when reaching $q$, 
whose delay then must be sufficient to satisfy all possible future inspection needs.
Accordingly, we define for every state $q$ of $A$, the set
\[
S[q] = \{\semsharp{t}_{A,q}\mid t\in\dom_B(\iota(q))\}
\]
of generalized inspection needs.
In order to compute this set, we construct a constraint system  $\C_\In$.
The system consists of one constraint per rule of $A$. 
Assume that $\tau$ is the rule $q(f(\ldots))\to p\{x_j\mapsto q_j(x_j)\mid j\in J\}$ where
$p\in\T_\Delta(X_J)$ and
$q_j,j\in J,$ are states of $A$. 
For $\tau$, the constraint system $\C_\In$ has the constraint
\[
\rho(q)\supseteq \{\semsharp{\tau}\{x_j\mapsto\alpha_j\mid j\in J\}\mid
	\forall j\in J.\,\alpha_j\in \rho(q_j)\}\,.
\]
The given constraint system is over subsets of $\In$, ordered by subset inclusion.
Since
all right-hand sides are monotonic in assignments $\rho:Q\to2^{\In}$,
the constraint system has a \emph{least} solution $\rho_0$.
Moreover, we have:

\begin{lemma}\label{l:fixpoint}
Assume that $A$ is a canonical earliest \DBTI\ transducer which is equivalent to some
\DBTI\ $A'$ without inspection needs. 
Let $\rho_0$ denote the least solution of the system of constraints $\C_\In$ for $A$. 
Then for each state $q$ of $A$:
\begin{enumerate}
\item $\rho_0(q) = S[q]$; and
\item the length of each $\alpha\in S[q]$ is bounded by $|A|$.
\end{enumerate}
\end{lemma}
\begin{proof}
In order to verify the first statement, we prove by induction that 
the $i$th iterate $\rho_0(q)^{(i)}$ of the constraint system exactly equals the
set of all $\semsharp{t}_{A,q}$ for trees $t$ of depth less than $i$.
For a proof of the second statement, we note 
that each $\alpha\in S[q]$ is the generalized inspection need of some execution starting
in $q$ which must be accomplished by every pattern available at $q$. 
%
%
%
If the axiom of $A$ is a ground tree $T_0$, then the set of states of $A$ is empty,
and the lemma trivially holds.

Otherwise, for every state $q$ of $A$ with $\iota(q) = h$, there is a context $t\in\C_\Sigma$ such that
$\sem{t}_{A}^h = u\cdot q(x_1)$ holds for some $u\in\T_\Sigma(x_1)$.
Then $t$ can be chosen in such a way that $u$ consists of at most
$|A|$ irreducible factors, i.e., $|u|\leq |A|$. 
%
%
Since at least one factor of the pattern $u$ is required to realize the inspection at one rule,
the upper bound on the lengths of inspection needs $\alpha\in S[q]$ follows.
\end{proof}

\noindent
As a consequence, the sets $S[q]$, $q$ a state of $A$, are effectively computable.
%
%

\subsection{The Construction}\label{ss:construction}

The input to the construction is a canonical earliest \emph{uc}-transducer $A$.
If the axiom of $A$ is ground, then the translation of $A$ can be realized
by a \DBTI\ $A'$ without inspection needs iff $\dom_B(h_0)$ is output-realizable by $T_0$.
Therefore, now assume that the axiom is of the form $u_0\cdot q_0(x_1)$ for some state $q_0$ of $A$.

For a finite set $S\subseteq\In$, and $s\in\T_\Sigma(x_1)$, let
$[s]^S$ denote the minimal suffix $v$ of $s$ such that $\semsharpsharp{v}\alpha=\epsilon$ 
for all $\alpha\in S$.
In a first attempt, we may try to construct $A'$ using the states of $A$ only.
This attempt, however, may fail as illustrated by the following example.

\begin{example}\label{e:buffer}
Consider the canonical earliest transducer with 
a single state $q$, the axiom $f_1(a,f_2(a,q(x_1)))$ and rules
\[
\begin{array}{lll}
q(f(x_1:h_0)) &\to& f_3(q(x_1),a)	\\
q(g(x_1:h_1,x_2:h_2,x_3:h_3)) &\to& a     
\end{array}
\]
where $\dom_B(h_1)$, $\dom_B(h_2)$, and $\dom_B(h_3)$ all are assumed to be output-recognizable by $a$.
Depending on the input tree, the suffixes of possibly delayed outputs to meet the inspection needs (beyond the ground right-hand side)
of the last rule are $f_1(a, f_2(a,x_1))$, $f_2(a,f_3(x_1,a))$, or $f_3(f_3(x_1,a),a)$.
Accordingly, the state $q$ must be split into three in order to deal with these three distinct situations.
\qed
\end{example}

\noindent
The states of the new \DBTI\ $A'$ therefore are \emph{pairs} $\angl{q,s}$, $q$ a state of $A$ 
and $s\in\T_\Delta(x_1)$ an output pattern for $A$, 
which we will call the \emph{buffer}.
Assume that $u_0 = u\cdot v$ where $v= [u_0]^{S[q_0]}$.
Then the axiom of $A'$ is given by $u\cdot\angl{q_0,v}(x_1)$.
Assume that state $\angl{q,u}$ of $A'$ has already been constructed, 
and $\tau$ is a rule of $A$ of the form
$
q(f(\ldots))\to p\{x_j\mapsto q_j(x_j)\mid j\in J\}
$
where $p\in\T_\Delta(X_J)$.
We make a case distinction on the cardinality of $J$.

\begin{itemize}
\item
If $J=\emptyset$,
i.e., $p\in\T_\Delta$, then $A'$ has a rule $\angl{q,u}(f(\ldots))\to u\cdot p$.
In case that $\eta_\tau\neq\emptyset$, $u\cdot p$ must be sufficient to 
satisfy the inspection needs incurred by the rule $\tau$, i.e.,
$\angl{\angl{\{u\cdot t\},\eta_\tau}}$ must contain $\eta_\tau$.
\item
Next assume that $J=\{j\}$, i.e.,
$p = p'\cdot x_j$.
Then $\semsharpsharp{u\cdot p'} ((\eta_\tau,\{\emptyset\})\,\alpha) = \epsilon$ 
must hold for all $\alpha\in S[q_j]$. 
Therefore, there is a factorization such that $u\cdot p' = u'\cdot v$
where $v= [u\cdot p']^{S[q_j]}$; and we add the rule
\[
\angl{q,u}(f(\ldots))\to u'\cdot\angl{q_j,v}(x_j) 
\]
to $A'$.
Accordingly, the assumption is that 
$\semsharpsharp{u'} (\eta_\tau,\{\emptyset\}) = \epsilon$ holds.
\item
Finally, assume that $J$ contains at least two elements. Then $p$ is of the form
$p = p'\{x_j\mapsto p_j\cdot q_j(x_j)\mid j\in J\}$ for $p'\in\T_\Delta(X_J)$
and maximal patterns $p_j\in\T_\Delta(x_1)$.
For each $j\in J$, there must be a factorization $p_j= u_j\cdot v_j$
where $v_j=[p_j]^{S[q_j]}$. Moreover, 
the multiset $G$ of ground subtrees of $u$, $p'$ and $u_j,j\in J$,
must be sufficient to satisfy the inspection need of $\tau$ itself, i.e., 
$\angl{\angl{G,\eta_\tau}}$ contains $\eta_\tau$. Then we add to $A'$ the rule
\[ 
\angl{q,u}(f(\ldots))\to u\cdot p'\{x_j\mapsto u_j\cdot \angl{q_j,v_j}(x_j)\mid j\in J\}.
\]
\end{itemize}

\begin{example}\label{e:buffer-1}
Consider again the \DBTI transducer from Example \ref{e:buffer}.
Let $a_1 = f_1(a,x_1), a_2 = f_2(a,x_1)$ and $a_3 = f_3(x_1,a)$ be the 
irreducible factors possibly occurring in the buffer.
Then the set of possible inspection needs for $q$ is given by
\[
\begin{array}{llll}
S[q] & = & \{	\\
	& & (\{(1,h_1),(2,h_2),(3,h_3)\},& \{\emptyset\}),	\\
	& & (\{(1,h_1),(2,h_2),(3,h_3)\},& \{\{(1,h_1)\},\{(2,h_2)\},\{(3,h_3)\},\emptyset\}),	\\
	& & (\{(1,h_1),(2,h_2),(3,h_3)\},& \{	
	      \{(1,h_1),(2,h_2)\},	\\
	& & & \phantom{\{}\{(1,h_1),(3,h_3)\}, \\
	& & & \phantom{\{}\{(2,h_2),(3,h_3)\},\\
	& & & \phantom{\{}
	\{(1,h_1)\},\{(2,h_2)\},\{(3,h_3)\},\emptyset\}),	\\
	& &\epsilon	\\
	& &\}\,.
	\end{array}
\]
The set of new states then consists of $\angl{q,a_1 a_2}, \angl{q,a_2 a_3}$, and $\angl{q,a_3 a_3}$
where the new axiom and rules are given by $\angl{q,a_1a_2}(x_1)$ and the following rules:
\[
\begin{array}{lll}
\angl{q,a_1 a_2}(f(x_1:h_0)) &\to& f_1(a,\angl{q,a_2 a_3}(x_1))	\\
\angl{q,a_2 a_3}(f(x_1:h_0)) &\to& f_2(a,\angl{q,a_3 a_3}(x_1))	\\
\angl{q,a_3 a_3}(f(x_1:h_0)) &\to& f_3(\angl{q,a_3 a_3}(x_1),a)	\\
\angl{q,a_1a_2}(g(x_1:h_1,x_2:h_2,x_3:h_3)) &\to& f_1(a,f_2(a,a))\\
\angl{q,a_2a_3}(g(x_1:h_1,x_2:h_2,x_3:h_3)) &\to& f_2(a,f_3(a,a))\\
\angl{q,a_3a_3}(g(x_1:h_1,x_2:h_2,x_3:h_3)) &\to& f_3(f_3(a,a),a).
\end{array}
\]
 \qed
\end{example}
Correctness and termination of the construction follows from the subsequent
Lemma~\ref{l:inspection_term}.

\begin{lemma}\label{l:inspection_term}
Assume that $A$ is the canonical earliest \DBTI\ for a 
\DBTI\ $A''$ without inspection needs.
%
Then the construction 
terminates with some $A'$ so that the following properties are satisfied:
\begin{enumerate}
\item	$A'$ is equivalent to $A$;
\item	Each inspection need $\eta_\tau$ of $A'$ is satisfiable;
\item	For each constructed state $\angl{q,u}$ of $A'$,
	$u\in\T_\Delta(x_1)$ has at most $(a-1)\cdot n\cdot|A|^2$ irreducible
	factors, each of size at most $|A|$, where $n$ is the number of states of $A$,
	and $a$ is the maximal rank of an input symbol.
	%
	%
\end{enumerate}
\end{lemma}

\begin{proof}
Assume that $A$ is equivalent to some $A''$ without inspection need and without constant states, 
i.e., is the canonical earliest normal form of $A''$.
This means that for 
states $q_1,q_2$ of $A$ with $\iota(q_2) = h$, 
$t_1\in\C_\Sigma$ and $t_2\in\T_\Sigma$ with 
$\sem{t_1}_{A,q_1}^{h} = s_1\cdot q_2(x_1)$, and $\sem{t_2}_{A,q_2} = s_2$  and 
all states $q'_1,q'_2$ of $A''$ so that $q'_i\in q_i$,
$\sem{t_1}_{A'',q'_1}^{h} = s'_1\cdot q'_2(x_1)$ and $\sem{t}_{A'',q'_2} = s'_2$,
for some $s'_1,s'_2$, it holds that 
$v_1\cdot s_1 = s'_1\cdot v_2$ and $v_2\cdot s_2= s'_2$ 
for $v_i = \pref_{A'}(q'_i)$.

Consider a pair $\angl{q,u}$ returned by the construction and let $q'\in q$
denote a state of $A''$ contained in $q$ with $v_{q'}=\pref_{A''}(q')$.
Then $u$ is a suffix of $v_{q'}$. Consequently, the set of 
all constructed states $\angl{q,u}$ is finite, and all pre-conditions at the
construction of rules are met. 
This means that all inspection needs of the resulting transducer are satisfiable.
Moreover, we have that
$u\cdot \sem{t}_{A,q} = \sem{t}_{A',\angl{q,u}}$ holds ---
implying that $A$ and $A'$ are equivalent.

It remains to prove item (3). Since by construction, we only put irreducible factors into
the buffer which occur as factors in unary variable output subtrees, their sizes are necessarily bounded by
$|A|$. 
%
%
Therefore, it remains to provide an upper bound for the lengths of the patterns $u$
occurring in 
constructed states $\angl{q,u}$ --- not in terms of the
transducer $A''$, but in terms of the transducer $A$, which serves as input to the construction.
According to our construction, $u$ is a minimal pattern to satisfy all generalized
inspection needs in $S[q]$.

We start with the following observation.
Consider contexts $t_0,t\in\C_\Sigma$ so that 
$\sem{t_0}_{A}^h = s_0\cdot q(x_1)$ and 
$\sem{t}_{A,q}^h = s\cdot q(x_1)$ 
holds for $h=\iota(q)$ and suitable $s\in\T_\Delta(x_1)$
where 
$\sem{t_0}_{A'}^h = s'_0\cdot\angl{q,u}(x_1)$ and
$\sem{t}_{A',\angl{q,u}}^h = s'\cdot\angl{q,u'}(x_1)$ where
$s_0 = s'_0\cdot u$ holds. Moreover by left cancellation,
\[
u\cdot s = s'\cdot u'
\]
holds since $s'_0\cdot u\cdot s = s_0\cdot s = s'_0\cdot s'\cdot u'$.
Moreover,
\[
\semsharp{t}_{A,q}^h(\alpha) \in S[q]\quad\text{for all}\quad \alpha\in S[q].
\]
If furthermore, $s\neq x_1$, then for every pair $(M,\phi)$ occurring in some generalized inspection need 
$\alpha\in S[q]$, there is some $M'\in\phi$ so that for every $(j,h)\in (M\setminus M')$, $\dom_A(h)$ is 
output-recognizable w.r.t.\ some distinct ground subtree of $s$.
In particular, if $\phi=\emptyset$ then every $(j,h)\in M$, $\dom_A(h)$ is
output-recognizable w.r.t.\ some distinct ground subtree of $s$.
%
%

Let $a$ denote the maximal arity of input symbols, and $m\leq|A|$ denote the maximal 
length of a generalized inspection need in $S[q]$,
and assume for $N\geq (a-1)\cdot m$ that there are contexts $t_1,\ldots,t_N$ all with
$\sem{t_i}_{A,q}^h = s_i\cdot q(x_1)$ with $s_i\neq x_1$. 
Then for $t = t_1\cdot\ldots\cdot t_N$, $\sem{t}_{A,q}^h = s\cdot q(x_1)$ with
$s = s_1\cdot\ldots\cdot s_N$ contains enough ground subtrees to satisfy all generalized inspection needs 
of $S[q]$.
This means that $\sem{t}_{A',\angl{q,u}}^h = s'\cdot\angl{q,u'}(x_1)$ with $u'$ 
necessarily being a suffix of $s$.
%
We use this observation to prove our claim.
Assume $\sem{t}_{A}^{h'} = s\cdot q'(x_1)$ and $\sem{t}_{A',\angl{q,u}}^{h'} = s'\cdot\angl{q',u'}(x_1)$
so that $s = s'\cdot u'$.
Assume for a contradiction, that $t$ is chosen minimal such that $|u'| > (a-1)\cdot n\cdot |A|^2 $.
Then there is some $N > (a-1)\cdot|A|$, and factorizations 
$t = t_0\cdot t_1\cdot\ldots\cdot t_N\cdot t_{N+1}$,
$s = s_0\cdot t_1\cdot\ldots\cdot s_N\cdot s_{N+1}$,
and some state $q$ with $\iota(q) = h$
such that
$\sem{t_0}_{A}^{h} = s_0\cdot q(x_1)$,
$\sem{t_i}_{A,q}^{h} = s_i\cdot q(x_1)$ with $s_i\neq x_1$ for $i=1,\ldots,N$, and
$\sem{t_{N+1}}_{A,q}^{h'} = s_{N+1}$ with
$|t_i|\leq (n-1)\cdot(a-1)\cdot|A|$ holds for all $i=1,\ldots,N+1$.
\ignore{
This means that also
$|s_i|\leq (n-1)\cdot(a-1)\cdot|A|^2$ holds for all $i=1,\ldots,N+1$.
}
By construction, already $s_2\cdot\ldots\cdot s_{N+1}$ is sufficient to satisfy
all inspection needs -- implying that $u'$ must be a suffix of 
of $\sem{t_0\cdot t_2\cdot\ldots\cdot t_{N+1}}_{A}^{h'}$ -- in contradiction to the minimality of $t$.
\ignore{
By construction, $s_1\cdot\ldots\cdot s_{N+1}$ is shorter than $u'$ and therefore, a proper suffix.
On the other hand, $s_1\cdot\ldots\cdot s_N$ suffices to satisfy all generalized inspection needs in $S[q]$ --
implying that $s_1\cdot\ldots\cdot s_{N+1}$ must satisfy all inspection needs of $S[q']$.
That means that $u'$ in fact also must be a suffix of $s_1\cdot\ldots\cdot s_{N+1}$ -- contradiction.
}
We conclude thus that the maximal length of the buffer is bounded by $(a-1)\cdot n\cdot|A|^2$.
\end{proof}

\noindent
In summary, we have shown:

\begin{theorem}\label{t:inspection}
\begin{enumerate}
\item	For a \DBTI\ (\DLTI) $A$, it is decidable whether or not $A$ is equivalent
	to some \DBT\ (\DLT) without inspection.
	%
\item	For a \DBTR\ (\DLTR) $A$ it is decidable whether or not $A$ is equivalent
	to some \DBT\ (\DLT) without look-ahead and without inspection.
	%
\end{enumerate}
\end{theorem}

\begin{proof}
By Theorem \ref{t:canonical}, we can w.l.o.g.\ assume that $A$ is canonical earliest.
By Lemma \ref{l:inspection_term}, it is decidable whether or not $A$ 
is equivalent to a \DBTI (\DLTI) without inspection needs, and if so, an equivalent \DBTI (\DLTI) $A'$ 
without inspection needs be can be constructed.

For any \DBTI (\DLTI) $A'$ without inspection needs, an equivalent \DBT (\DLT) $A''$ without inspection can be constructed.
Likewise, for each 
\DBT (\DLT) $A''$ without inspection, an equivalent \DBTI (\DLTI) $A'$ with inspection, but without inspection needs 
can be constructed.
From that, statement (1) follows.

Statement (2) then follows from statement (1) together with Theorem \ref{t:removal}. 
\end{proof}
\ignore{
\noindent
It should be clear that analogous constructions also can be applied in case that
the transducers are \emph{linear}. We therefore obtain:

\begin{theorem}\label{t:inspection_L}
\begin{enumerate}
\item	For an \DLTI\ $A$, it is decidable whether or not $A$ is equivalent
	to some \DLT\ without inspection.
	%
\item	For an \DLTR\ $A$ it is decidable whether or not $A$ is equivalent
	to some \DLT\ without look-ahead and without inspection.
	%
	\qed
\end{enumerate}
\end{theorem}
}

\noindent
In Section~\ref{s:basics} we have seen that a bottom-up deterministic transducer
(\DB)
can be seen as a particularly simple \DBTR\ transducer. 
Similarly, a linear \DB\ transducer can be considered as a \DLTR\ transducer.
Accordingly, we obtain as a corollary:

\begin{corollary}
Let $A$ be a (linear) \DB\ transducer.
It is decidable if an equivalent \DBTI (\DLTI) or an equivalent
\DBT (\DLT) transducer exists, and if so, such a transducer can be constructed.
\qed
\end{corollary}

\section{Conclusion}\label{s:conclusion}

We showed for two natural subclasses of deterministic top-down tree transducers
(namely, \emph{linear} and \emph{uniform-copying} ones)
how to remove
(bottom-up deterministic) look-ahead and 
replace it whenever possible, with (top-down deterministic) inspection. 
We then also showed for the given classes how to remove inspection (if possible).
The constructions are technically intricate, but crucially rely on canonical earliest
normal forms for the transducers in question. 
As a corollary we obtain that for a given deterministic bottom-up transducer it is
decidable whether or not it can be realized by a deterministic top-down tree transducer
that is either\emph{linear} or \emph{uniform-copying}.
Note that, using results of this paper (or more precisely its predecessor~\cite{DBLP:conf/icalp/ManethS20})
it was shown recently that for an arbitrary deterministic top-down tree transducer
with look-ahead it is decidable whether or not it can be realized by a 
\emph{linear} transducer~\cite{DBLP:journals/ijfcs/ManethSV23}.

One may wonder if our results imply that for a given deterministic bottom-up tree 
transducer $U$ it is decidable whether or not it can be realized by an
\emph{arbitrary} deterministic top-down tree transducer, 
i.e., if $U$ can be realized by a top-down transducer, can it also be realized by
a \emph{uniform-copying}
such transducer?
Interestingly, this is \emph{not} necessarily the case:
let $h_a, h_b$ be look-ahead states that indicate that the left-most
leaf of the input tree is labeled $a$ or $b$, respectively, and consider
a transducer $U$ which has rules (for every $h\in\{h_a,h_b\}$):
\[
\begin{array}{l@{\qquad}l}
q_0(f(x_1:h_a,x_2:h))\to g(a, b, q_{\text{id}}(x_2)) \\
q_0(f(x_1:h_b,x_2:h))\to g(c, d, q_{\text{id}}(x_2))
\end{array}
\]
The corresponding translation \emph{can} be realized by a deterministic
top-down tree transducer! However, there is no uniform-copying top-down tree transducer that 
realizes the translation realized by $U$: viz.,
the output leaves $a$ and $b$ must be produced by different states, but
both on the input variable $x_1$.
It remains open
whether the methods of this paper can be generalized so that the existence of equivalent 
possibly \emph{non-uniform-copying} transducers without look-ahead can be decided.
The results reported here all refer to the question inhowfar look-ahead can be removed altogether.
As a natural extension, however, is to ask whether a smaller look-ahead automation suffices to
realize a given translation or, whether a unique \emph{minimal} such automaton exists.

\ignore{
One may also wonder about precise complexity bounds for the decision procedures and 
constructions which we have provided. 
}

\ignore{
\section{Example Section}
\label{sec1}

Section text. See Subsection \ref{subsec1}.

\subsection{Example Subsection}
\label{subsec1}

Subsection text.


\subsubsection{Mathematics}
This is an example for the symbol $\alpha$ tagged as inline mathematics.

\begin{equation}
f(x) = (x+a)(x+b)
\end{equation}

\begin{equation*}
f(x) = (x+a)(x+b)
\end{equation*}

\begin{align}
 f(x) &= (x+a)(x+b) \\
      &= x^2 + (a+b)x + ab
\end{align}

\begin{eqnarray}
 f(x) &=& (x+a)(x+b) \nonumber\\ 
      &=& x^2 + (a+b)x + ab
\end{eqnarray}

\begin{align*}
 f(x) &= (x+a)(x+b) \\
      &= x^2 + (a+b)x + ab
\end{align*}

\begin{eqnarray*}
 f(x)&=& (x+a)(x+b) \\
     &=& x^2 + (a+b)x + ab
\end{eqnarray*}


\begin{table}[t]
\centering
\begin{tabular}{l c r}
  1 & 2 & 3 \\ 
  4 & 5 & 6 \\
  7 & 8 & 9 \\
\end{tabular}
\caption{Table Caption}\label{fig1}
\end{table}

\begin{figure}[t]
\centering
\includegraphics{example-image-a}
\caption{Figure Caption}\label{fig1}
\end{figure}

\appendix
\section{Example Appendix Section}
\label{app1}

Appendix text.


Example citation, See \cite{lamport94}.
}





\end{document}